\documentclass[preprint,3p,onecolumn]{elsarticle}

\usepackage{graphicx}
\usepackage{bm}
\usepackage{threeparttable}
\usepackage{booktabs}
\usepackage{commath}
\usepackage{array}
\usepackage{enumerate}
\usepackage{paralist}
\usepackage{amsmath,amssymb}
\usepackage[colorlinks]{hyperref}
\usepackage[utf8]{inputenc}
\usepackage{natbib}
\usepackage[capitalize]{cleveref}
\crefname{section}{Sec.}{Secs.}
\Crefname{section}{Section}{Sections}
\crefname{equation}{Eq.}{Eqs.}
\Crefname{equation}{Equation}{Equations}
\crefname{figure}{Fig.}{Figs.}
\Crefname{figure}{Figure}{Figures}
\crefname{subsection}{Sec.}{Secs.}
\usepackage{graphicx}
\usepackage{float}
\usepackage{verbatim}
\usepackage{cuted}
\usepackage{stfloats}
\usepackage[normalem]{ulem}
\usepackage{subcaption}
\usepackage{stfloats}
\usepackage{appendix}
\usepackage{xcolor}

\newcommand{\reviewer}[1]{#1}

\newcommand{\cA}{\mathcal{A}}

\newcommand{\fu}{\mathfrak{u}}

\newcommand{\gE}{\gamma_{\rm E}}

\newcommand{\ii}{\mathrm{i}}

\journal{Physics Letters B}

\begin{document}

\begin{frontmatter}


\title{Simultaneous bounds on the gravitational dipole radiation and varying
gravitational constant from compact binary inspirals}



\author[1,2]{Ziming Wang}
\author[3]{Junjie Zhao\fnref{coauthor}}
\author[4,5]{Zihe An}
\author[2,6]{Lijing Shao\corref{cor1}}\ead{lshao@pku.edu.cn}
\author[3,7]{Zhoujian Cao}
\fntext[coauthor]{The first two authors contributed equally to the work.}
\cortext[cor1]{Corresponding author}
%--
\address[1]{Department of Astronomy, School of Physics, Peking University,
Beijing 100871, China}
\address[2]{Kavli Institute for Astronomy and Astrophysics, Peking University,
Beijing 100871, China}
\address[3]{Department of Astronomy, Beijing Normal University, Beijing 100875,
China}
\address[4]{Department of Physics, Tsinghua University, Beijing 100084, China}
\address[5]{Department of Mechanics and Engineering Science, 
College of Engineering, Peking University, Beijing 100871, China}
\address[6]{National Astronomical Observatories, Chinese Academy of Sciences,
Beijing 100012, China}
\address[7]{School of Fundamental Physics and Mathematical Sciences, Hangzhou Institute for Advanced Study, UCAS, Hangzhou 310024, China}
%--

\begin{abstract}
Compact binaries are an important class of gravitational-wave (GW) sources that
can be detected by current and future GW observatories. They provide a testbed
for general relativity (GR) in the highly dynamical strong-field regime.  Here,
we use GWs from inspiraling binary neutron stars and binary black holes to
investigate dipolar gravitational radiation (DGR) and varying gravitational
constant predicted by some alternative theories to GR, such as the scalar-tensor
gravity. Within the parametrized post-Einsteinian framework, we introduce the
parametrization of these two effects simultaneously into compact binaries'
inspiral waveform and perform the Fisher-information-matrix analysis to estimate
their simultaneous bounds. \reviewer{In general, the space-based GW detectors
can give a tighter limit than ground-based ones. The tightest constraints can
reach $\sigma_B<3\times10^{-11}$ for the DGR parameter $B$ and
$\sigma_{\dot{G}}/G < 7\times10^{-9} \, {\rm yr}^{-1} $ for the varying $G$,
when the time to coalescence of the GW event is close to the lifetime of
space-based detectors.} In addition, we analyze the correlation between these
two effects and highlight the importance of considering both effects in order to
arrive at more realistic results.
\end{abstract}

\end{frontmatter}

% \linenumbers
% \reviewer{use this for replies to Reviewer \#1}
%---------------------------------------------------------------------
\section{Introduction}
\label{sec:intro}
%---------------------------------------------------------------------

Until now, Einstein's theory of general relativity (GR)~\cite{Einstein:1915ca}
is a widely accepted theory passing a bulk of precise experimental tests with
flying colors~\cite{Will:2018bme}. Nevertheless, there still exist a vast amount
of motivations to explore alternative theories of gravity. Theoretically, the
incompatibility between GR and quantum mechanics remains to be solved. From the
experimental point of view, GR hardly explains the observations like the cosmic
inflation and anomalous kinematics of galaxies without introducing the unknown
dark energy or dark matter~\cite{Berti:2015itd}. GR is still an imperfect theory
of gravity to a certain extent. As an effort to go beyond GR, various extensions
and modifications to GR were proposed~\cite{Berti:2015itd}.  Among possible
modifications to GR from numerous perspectives, we consider time variation of
the gravitational constant $G$ and the dipolar gravitational radiation (DGR) in
this work. These two aspects represent some popular strawman targets for
phenomenological studies of gravity~\cite{Will:2018bme}.

In GR, the gravitational coupling strength $G$ is a constant independent of
space and time.  However, this cannot be taken for granted.
\citet{Dirac:1937ti,Dirac:1973zw} first came up with the possibility of a
time-varying $G$, hinted by his large number hypothesis. Theoretically, some
alternative theories of gravity indicate a varying gravitational constant, 
notably in scalar-tensor theories (STTs)~\cite{Brans:1961sx,Fujii:2003pa},
higher-dimensional theories~\cite{Marciano:1983wy,Kolb:1985sj,Landau:2000cc},
and string theories~\cite{Maeda:1987ku,Taylor:1988nw}. There are lots of
experiments for testing gravity with the time-varying $G$ effect, such as lunar
laser ranging experiments~\cite{Williams:1976zz, Williams:1995nq,
Williams:2004qba}, radio observations of binary
pulsars~\cite{Nordtvedt:1990zz, Damour:1988zz, Shao:2016ezh, Miao:2021awa},
cosmic microwave background observations~\cite{Wu:2009zb}, and Big Bang
nucleosynthesis~\cite{Bambi:2005fi,Copi:2003xd}.

\reviewer{ A time-varying $G$ in most alternative theories is often accompanied
by the existence of one or more extra degrees of freedom in the gravitational
sector.  For instance, the Jordan-Fierz-Brans-Dicke (JFBD)
theory~\cite{Brans:1961sx,Jordan:1949zz, Jordan:1959eg, Fierz:1956zz}---one of
the simplest natural alternative STTs---involves an additional scalar field
$\phi$, which indicates a varying $G$ via $G \propto {\phi}^{-1}$ when $\phi$ is
time dependent~\cite{Will:1994fb}. In addition, additional gravitational degrees
of freedom can usually cause extra DGR in these alternative theories of gravity
\cite{Will:1994fb, Seymour:2018bce, Barausse:2016eii, Shao:2017gwu}, while
neither monopole nor dipole radiation exists in GR. DGR is characterized by a
parameter $B$ which vanishes in GR~\cite{Barausse:2016eii}.  As an extra channel
of energy dissipation, DGR enhances the orbital decay in a binary system.
Therefore, in some alternative theories it is advised to investigate these two
non-GR corrections simultaneously~\cite{Lazaridis:2009kq,
Zhu:2018etc,Wex:2014nva}.} 

In the weak-field regime, bounds have been put which can be associated to
DGR~\cite{Will:2014kxa}.  However, the weak-field experiments are not always
applicable to the strong-field regime~\cite{Damour:1993hw}. In this case, one
needs to investigate the motion of strongly self-gravitating bodies, such as
neutron stars (NSs) and black holes (BHs), which can more easily deviate from
the GR expectation in some modified gravity theories \cite{Barausse:2016eii}.

To explore DGR, binary pulsars in the quasi-stationary strong-field regime turn
out to be ideal testbeds~\cite{Shao:2016ezh, Wex:2014nva, Damour:2007uf,
Will:1977zz, Will:1977wq, Gerard:2001fm}. The emitted radio pulses from binary
pulsars can be continuously monitored by large radio telescopes. With the
technology called pulsar timing, the parameters of binary pulsars can be
measured to high precision and be used to detect or constrain the varying-$G$
and DGR effects simultaneously, as was done in
Refs.~\cite{Lazaridis:2009kq,Zhu:2018etc, Freire:2012mg}. One of the best
results came from a combination of PSRs J0437$-$4715, J1713$+$0747, and
J1738$+$0333~\cite{Zhu:2018etc}, which simultaneously constrained $\sigma_{\Dot
G}/G < 10^{-12} \, {\rm yr}^{-1}$ and $\sigma_{B} < 10^{-7}$ at the 68\%
confidence level (CL). 

Pulsars can put tight limits on varying-$G$ and DGR effects in several
gravitational theories, including the JFBD theory. In JFBD theory, binary NS
(BNS) inspirals can emit DGR while binary BHs (BBHs) cannot, because of the
no-hair theorem for BHs~\cite{Yunes:2011aa, Mirshekari:2013vb}. However, there
also exist other STTs which predict DGR predominantly (or only) in BBHs, e.g.\
the shift-symmetric Horndeski theories~\cite{Barausse:2015wia}. In those STTs
where DGR only comes from systems involving BHs, pulsar observations cannot put
limits, and we have to resort to other means, e.g., via gravitational-wave (GW)
observations.

Up to now, about a hundred GWs have been announced by the LIGO/Virgo/KAGRA
Collaboration from their observing runs O1, O2 and
O3~\cite{LIGOScientific:2018mvr, LIGOScientific:2020ibl,
LIGOScientific:2021djp}, including the first BBH event
GW150914~\cite{LIGOScientific:2016aoc} and the first BNS  event
GW170817~\cite{LIGOScientific:2017vwq}. A large number of events are expected in
the near future~\cite{Abbott:2020qfu}. GW signals are detected by GW
interferometers, and analysis based on the matched-filtering method recovers the
information about the dynamics of its source~\cite{Finn:1992wt}. They can also
be used to put constraints on the non-GR effects~\cite{Tahura:2018zuq,
Abbott:2018lct, Zhao:2019suc, Guo:2021leu, Liu:2020nwz, LIGOScientific:2021sio}.
Currently, all observed GWs come from compact binary coalescences in the highly
dynamical strong-field regime, providing another powerful testbed besides binary
pulsars. Some earlier studies~\cite{Abbott:2018lct, Yunes:2010qb, Zhao:2022vig}
have already used binary pulsars and GWs to constrain the DGR  and the
varying-$G$ effects.  However, these works usually focus on one aspect, and do
not include varying $G$ and DGR simultaneously. 

The corresponding phase corrections due to the varying-$G$  and DGR effects are
at the $-4$ post-Newtonian (PN) and $-1$\,PN orders respectively.\footnote{The
contribution of the $n$\,PN correction to GW phase is at the relative order of
$\sim (v/c)^{2n}$, where $v$ is the typical binary velocity.} In general,
relative to binary pulsars, BNS inspirals have a larger binary velocity $v$,
leading to weaker constraints. The bounds from BBH systems are usually several
orders of magnitude looser than those from BNSs for the ground-based GW
detectors, while for space-based GW detectors there is no such big difference.
This is because that BBH inspirals stay in the ground-based detectors' frequency
bands for a shorter time relative to BNS inspirals.  Similarly, in general,
space-based detectors can give tighter limits than ground-based detectors for
negative PN effects, because of their higher sensitivities in the low-frequency
band~\cite{Tahura:2018zuq}. 

In this work, we use the Fisher information matrix (FIM)~\cite{Finn:1992wt} to
investigate the prospects of testing the varying-$G$ and DGR effects
simultaneously with GW observations of BNSs and BBHs. We consider both
ground-based and space-based GW detectors, including the Advanced Laser
Interferometer Gravitational-Wave Observatory (AdvLIGO)~\cite{Abbott:2020qfu},
the third-generation GW detectors like the Einstein Telescope
(ET)~\cite{Hild:2010id, Sathyaprakash:2012jk, Maggiore:2019uih}  and Cosmic Explorer
(CE)~\cite{Evans:2016mbw,Reitze:2019iox,Reitze:2019dyk}, the proposed
space-based decihertz GW detectors like the Decihertz Observatory
(DO)~\cite{Sedda:2019uro, Sedda:2021yhn} and the DECi-hertz Interferometer
Gravitational wave Observatory (DECIGO)~\cite{Sato:2017dkf, Isoyama:2018rjb}
(see \cref{tab:curves}).\footnote{For space-based detectors, we use conservative
configurations ``B-DECIGO'' for DECIGO and ``DO-Conservative'' for DO. The ``B''
in B-DECIGO means the baseline sensitivity of DECIGO \cite{
Isoyama:2018rjb,Yagi:2011wg}, and ``DO-Conservative'' is a conservative
estimation of DO's sensitivity \cite{Sedda:2019uro, Sedda:2021yhn}.} For compact
binaries, we use GW150914 and GW170817 as examples of BBH and BNS events
respectively.\footnote{We use component masses $(m_1, m_2) = (35.6 \, M_{\odot}, 30.6
\, M_{\odot})$ for GW150914-like events and $(m_1, m_2) = (1.26 \, M_{\odot},
1.48 \, M_{\odot})$ for GW170817-like events.} 

From our study, for BNSs B-DECIGO can give the tightest constraints on DGR for
GW170817-like events, being four orders of magnitude tighter than binary
pulsars~\cite{Zhu:2018etc}. However, the constraints from AdvLIGO are looser
than those from binary pulsars. As for $\dot{G}$, the tightest bound from
GW170817-like events is still about four orders of magnitude looser than binary
pulsars~\cite{Zhu:2018etc}. These are in accordance with the above discussions.
We emphasize that, for some alternative gravity theories, only simultaneous
bounds on the varying-$G$ and DGR effects using GW events are creditable because
of the strong correlation between them. The absolute correlation coefficient
between them can reach $\sim 0.8$ or higher, and individual bounds on these effects
are only $15\%$--$40\%$ of the simultaneous bounds. The constraints on DGR
without considering the varying-$G$ effect are thus over-optimistic for some
specific alternative theories and vice versa.

We organize the paper as follows.  We summarize the main calculation method,
namely the FIM, in~\cref{sec:fisher}. In \cref{sec:gw:bns} we introduce briefly
the modified waveform template of GW inspiral signals of compact binaries and
the GW detectors we use. In~\cref{sec:bounds} we use the FIM to obtain joint
constraints on the varying-$G$ and DGR phenomena from simulated BNS and BBH GW
events. Conclusions are summarized in~\cref{sec:dis}. Throughout the paper, we
use the geometrized unit where $G=c=1$.

%---------------------------------------------------------------------
\section{Fisher information matrix}
\label{sec:fisher}
%---------------------------------------------------------------------

The matched-filtering analysis is the prescription used for parameter estimation
of GWs. We briefly review its general procedure here.  Firstly, the
noise-weighted inner product of two signals $h(t)$ and $g(t)$ is defined
by~\cite{Finn:1992wt}
\begin{equation}
(h\,, g)=4 \Re \int_{f_{\rm min}}^{f_{\rm max}} \frac{\tilde{h}(f)^{*}
\tilde{g}(f)}{S_{n}(f)} {\rm d} f \,, \label{eqn:product}
\end{equation}
where $\tilde{h}(f)$ and $\tilde{g}(f)$ are Fourier transformations of signals
$h(t)$ and $g(t)$ respectively.  In~\cref{eqn:product}, $S_n(f)$ is the
one-sided power spectral density (PSD) of the noise. The signal-to-noise ratio
(SNR) of a signal $h$ in the detector is defined as $\rho =(h, h)^{1/2}$.
Characterizing the strength of the signal, SNR is an important criterion to
assess whether we have detected a real GW event from the strain data. In
general, we set the critical SNR to $\rho = 8$ for ground-based GW detectors,
and $\rho = 15$ for space-based detectors \cite{Moore:2019pke, Liu:2021dcr}.

FIM is a useful tool for estimating the measurement uncertainties of GW
events~\cite{Finn:1992wt, Vallisneri:2007ev}. It is defined as,
\begin{equation}
    \label{eqn:fisher}
    F_{i j}=\left(\frac{\partial h}{\partial \theta_i}\,, \frac{\partial
    h}{\partial \theta_j}\right) \,,
\end{equation}
where $\theta_i$ and $\theta_j$ are the $i$-th and $j$-th parameters of the GW
waveform template $h$. Assuming a uniform prior, the inverse of the FIM yields
the covariance between the parameters in the posterior distribution, which is
approximately Gaussian in the high SNR limit or equivalently in the linear
signal approximation (LSA)~\cite{Vallisneri:2007ev, Wang:2022kia}. This
approximation of the full posterior can be proved by expanding the detected
signal around the true parameters up to the first order and substituting the
expansion into the Bayesian posterior~\cite{Vallisneri:2007ev}. Defining the
inverse of FIM as $\Sigma \equiv F^{-1}$, one can obtain the standard deviations
and correlation coefficients of the parameters 
%--
\begin{align}
    \label{eqn:FIMsigma}
    \sigma_i &= \sqrt{\Sigma_{i i}} \,, \\
    c_{i j} &= \frac{\Sigma_{i j}}{\sigma_i \sigma_j} =  \frac{\Sigma_{i
    j}}{\sqrt{\Sigma_{ii} \Sigma_{jj}}}\,, 
\end{align}
%--
where $\sigma_i$ is the standard deviation of the $i$-th parameter, and $c_{i
j}$ is the correlation coefficient between the $i$-th and $j$-th parameters
\cite{Finn:1992wt,Vallisneri:2007ev}. 

%---------------------------------------------------------------------
\section{Compact binary signals and GW observatories}
\label{sec:gw:bns}
%---------------------------------------------------------------------

%---------------------------------------------------------------------
% \subsection{Compact binary signals}\label{sec3.1:Compact Binary Signals}
%---------------------------------------------------------------------

Compact binary coalescences are highly dynamical gravitational systems whose
evolutions are described by inspiral, merger, and ringdown stages. The GW
waveform we use is from the inspiral stage assuming a quasi-circular orbit with
a quasi-adiabatic inspiral due to the emission of GWs~\cite{Will:1994fb,
Blanchet:2013haa, Buonanno:2009zt}.  Generally, the GW waveform in the Fourier
domain $\tilde{h}(f)$ is written as a function of the GW frequency $f$ in the
following restricted form
\begin{equation}
    \tilde{h}(f) = \cA f^{-7/6} e^{\ii \Psi(f)} \,,
    \label{eqn:waveform}
\end{equation}
where $\cA$ and $\Psi(f)$ stand for the amplitude and the phase, respectively.

%GR
In GR, the GW amplitude of a compact binary inspiral signal at the leading order
(i.e.\ the Newtonian order) is
\begin{equation}
    \cA_{\rm{GR}}=\frac{2}{5} \times \sqrt{\frac{5}{24}} \pi^{-2 / 3}
    \frac{\mathcal{M}^{5 / 6}}{d_L} \,,
    \label{eqn:amplitude}
\end{equation}
with $d_{L}$ the luminosity distance, $\mathcal{M} \equiv \eta^{3/5} M_z$ the
redshifted chirp mass, where $M_z \equiv (1+z) M$ with $M$ the sum of the
source-frame component masses $m_1$ and $m_2$, and $\eta = m_1 m_2/M^2$ is the
symmetric mass ratio. This amplitude has been averaged over the sky position and
polarization of GWs as well as the orientation of the
binary~\cite{Liu:2020nwz,Isoyama:2018rjb}, leading to the factor ``$2/5$'' in
the place of a function dependent on those angles.

Cutting at $3.5$\,PN, the inspiral phase of the binary in GR
reads~\cite{Buonanno:2009zt,Cutler:1994ys}
%--
\begin{align} 
\label{eqn:phase_GR}
    \Psi_{\rm{GR}}(f)=
    & 2 \pi f t_{c} - \Phi_{c} - \frac{\pi}{4} + \frac{3}{128 \eta} \fu^{-5
    / 3} \left\{ 1 + \left(\frac{3715}{756}+\frac{55}{9} \eta\right) \fu^{2
    / 3}-16 \pi \fu \right. \nonumber \\ 
    &+\left(\frac{15293365}{508032}+\frac{27145}{504} \eta+\frac{3085}{72}
    \eta^{2}\right) \fu^{4 / 3}+\pi\left(\frac{38645}{756}-\frac{65}{9}
    \eta\right)(1+\ln \fu) \fu^{5 / 3} \nonumber \\ 
    &+\left[\frac{11583231236531}{4694215680}-\frac{640}{3}
    \pi^{2}-\frac{6848}{21} \gE -\frac{6848}{63} \ln (64
    \fu)+\left(-\frac{15737765635}{3048192}+\frac{2255}{12} \pi^{2}\right)
    \eta\right. \nonumber \\ 
    &\left.\left. +\frac{76055}{1728} \eta^{2} -\frac{127825}{1296}
    \eta^{3}\right]
    \fu^{2}+\pi\left(\frac{77096675}{254016}+\frac{378515}{1512}
    \eta-\frac{74045}{756} \eta^{2}\right) \fu^{7 / 3}\right\} \,,
\end{align}
%--
where $\fu \equiv \pi M_z f$, $t_{c}$ and $\Phi_{c}$ are time and phase at
coalescence, $\gE$ is the Euler constant. Here, we neglect the spin effects in
the GW waveform. \Cref{eqn:waveform} with~\cref{eqn:amplitude,eqn:phase_GR}
gives the so-called ``restricted TaylorF2 waveform''~\cite{Buonanno:2009zt}
characterizing the inspiral signal in GR. It is valid until the end of the
inspiral when the binary comes to the innermost stable circular orbit (ISCO) and
plunges into a merger. Thus, the $f_{\max}$ in the inner product of
\cref{eqn:product} is chosen to be $f_{\max} = \min \left\{f_{\rm upper}, f_{\rm
ISCO}\right\}$, where $f_{\rm upper}$ and $f_{\rm ISCO}$ are the upper limit of
the detector's frequency range and the GW frequency corresponding to the ISCO
respectively.  In different alternative theories of gravity, the evolution of
BNSs is modified, leaving non-GR imprints in the amplitude and the phase of GWs.

%G
The varying $G$ makes a difference to the orbital dynamics in various aspects.
First of all, time variations of $G$ cause corrections in orbital frequency
since $f_{\rm orb} \simeq \sqrt{G M/{4 \pi^2\,r^3}}$ in the Newtonian
approximation, where $r$ is the orbital separation. Secondly, the decay of the
orbital frequency is also modified, as the amount of energy being lost from
binary systems due to emission of GWs is different~\cite{Seymour:2018bce}. In
addition, there can be time variations of component masses along with their
sensitivities defined by $s=-\partial \ln m/{\partial \ln
G}$~\cite{Will:1994fb}, as well as non-conservation of the binding energy due to
an acceleration caused by time variations in $G$ or
masses~\cite{Tahura:2018zuq}.

To include the varying-$G$ effect in the waveform, the following GW phase
modification corresponding to a $-4$\,PN term is added to the
phase~\cite{Tahura:2018zuq},
%--
\begin{equation}
    \label{eqn:phase_-4}
    \Psi_{\rm{-4\,PN}}(f)= - \frac{25}{851968} \frac{\dot G
    \mathcal{M}}{\eta^{13/5}} \left(11 - \frac{35}{2} S - \frac{41}{2} \sqrt{1 -
    4 \eta} \Delta S \right) \fu^{-13/3}\,, 
\end{equation}
%--
where $S \equiv s_1 + s_2$ and
$\Delta S \equiv s_1 - s_2$ with the component sensitivities $s_1$ and $s_2$. For a NS, the sensitivity $s$ is dependent on its mass
and equation of state, and a reasonable approximation is~\cite{Zhu:2018etc}
\begin{equation}
    s_i \simeq \frac{0.16}{1.33 M_{\odot}} m_i \,.
    \label{eqn:sensitivity}
\end{equation}
For a BH, we fix its sensitivity to $s = 0.5$ \cite{Berti:2018cxi}. 

%B
The DGR affects the binary evolution by shedding additional energy and momentum,
which leads to a faster orbital decay rate  and an earlier
coalescence~\cite{Barausse:2016eii}. The strength of DGR is parameterized by $B$
characterizing the extra energy loss due to DGR, which is defined
via~\cite{Barausse:2016eii},  
%%--
\begin{equation}
    \label{eqn:flux}
    \dot{E}_{\rm{GW}} =
    \dot{E}_{\rm{GR}}\left[1+B\left(\frac{M}{r}\right)^{-1}\right]\,.
\end{equation}
%%-- 
For example, in JFBD-like theory, $ B=5(\Delta \alpha)^{2}/96$, where $\Delta
\alpha$ is the difference of the scalar couplings of the two bodies. While the
determination of $B$ depends on specific gravity theory, the DGR modification to
the waveform is general, which reads,
\begin{equation}
    \label{eqn:phase_-1}
    \Psi_{\rm{-1\,PN}}(f)= - \frac{3}{224\eta} B \fu^{-7 / 3} \,.
\end{equation}
%

%amplitude
As for the amplitude, the modifications from DGR and varying-$G$ effects are
subdominants, which can be ignored. The phase plays a more important role in GW
analysis than amplitude in the matched-filtering method. Therefore, in this work
we choose
\begin{equation}
    \cA=\cA_{\rm{GR}} \,,
\end{equation}
\begin{equation}    
    \Psi(f)=\Psi_{\rm{GR}}(f)+\Psi_{\rm{-1\,PN}}(f)+\Psi_{\rm{-4\,PN}}(f) \,,
\end{equation}
as the waveform template for the inspiral signals. 

%---------------------------------------------------------------------
% \subsection{GW observatories}
%---------------------------------------------------------------------

\reviewer{We consider the following two types of GW detectors: (i) ground-based
laser interferometers, including AdvLIGO and Voyager, as well as the
third-generation ones like ET \cite{Hild:2010id} and CE~\cite{Evans:2016mbw,
Reitze:2019iox, Reitze:2019dyk}; and (ii) proposed space-based decihertz GW
detectors, DO-Conservative \cite{Sedda:2019uro} and B-DECIGO \cite{Sato:2017dkf,
Isoyama:2018rjb}.\footnote{We do not consider millihertz GW detectors in this
work, as they are less sensitive to GW signals from merging stellar-mass compact
objects~\cite{Liu:2021dcr}.} The AdvLIGO, together with the advanced Virgo and
KAGRA, has discovered about a hundred GW events up to now
\cite{LIGOScientific:2018mvr, LIGOScientific:2020ibl, LIGOScientific:2021djp}.
Voyager, ET and CE are proposed detectors for observing GWs extending the
AdvLIGO's frequency range with higher sensitivities, while DO-Conservative and
B-DECIGO are proposed to observe GWs in the decihertz frequency
band.} With the improvement of sensitivity, complimentary detection bands, and
an increase in the number of detectors, more GW events from a wider range of
source populations are expected to be detected. Note that a lower accessible
frequency means a longer observational time, whereas the observation time will
be limited by the lifetimes of the detectors. This will lead changes to the
$f_{\min}$ parameter in the integral range of the inner product of
\cref{eqn:product}. We choose the lifetime as 4 years for space-based detectors.
Therefore, the $f_{\min}$ in \cref{eqn:product} is taken as $f_{\min} = \max
\left\{f_{\rm lower}, f_{\rm 4\,yr}\right\}$, where $f_{\rm lower}$ is the low
limit of the detector's frequency range and $f_{\rm 4\,yr}$ is determined by 
%--
\begin{equation}
    f_{\rm 4\,yr} =\left({f_{\rm max}^{-8/{3}}} +\frac{256 \pi^{8/3} }{5} {\cal
    M}^{5/3} \cdot\Delta t_{\rm obs }\right)^{-3/8}\,,
\end{equation}
where $\Delta t_{\rm obs} = {\rm 4\, yr}$, and $f_{\rm max}$ is defined above.
\reviewer{As for the ground-based detectors, we choose $f_{\min} = f_{\rm
lower}$. }

\begin{figure*}[ht]
    \centering 
    \includegraphics[width=10cm]{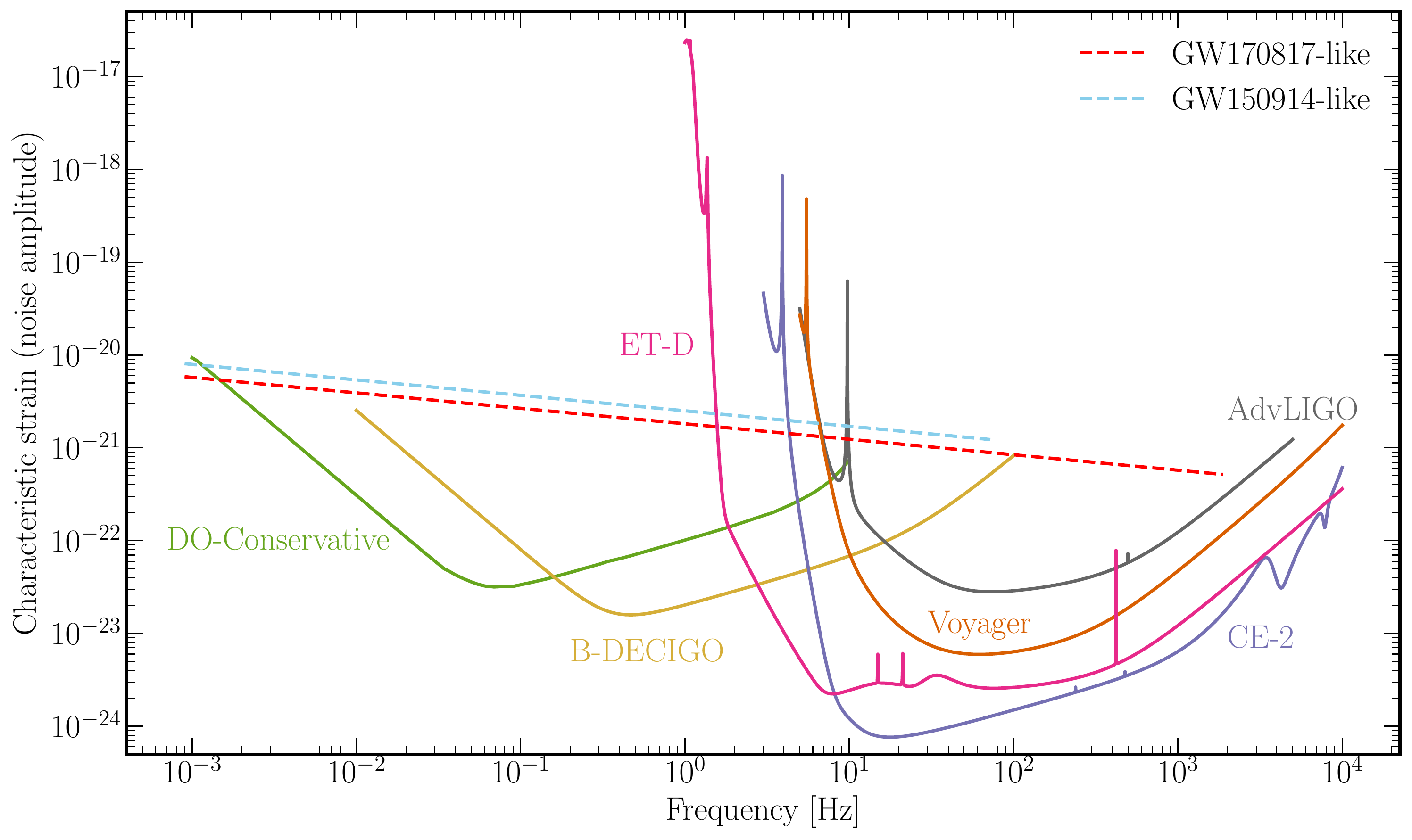}
    \caption{The dimensionless pattern-averaged characteristic inspiral strains
    of GW150914-like and GW170817-like events, in contrast to the dimensionless
    noise spectral density of the detectors. The characteristic strains are
    averaged over sky location and polarization angle of the source, ignoring
    the spin contribution and truncating at $f_{\rm {ISCO}}$. The noise  is
    plotted according to the configurations listed in~\cref{tab:curves}. }
    \label{fig:characteristic strain}
\end{figure*}
%%

%---------------------------------------------------------------------
\begin{table}
    \centering
    \renewcommand{\arraystretch}{1.25}
    \caption{Characteristics of GW detectors. \label{tab:curves}}
    \begin{threeparttable}
    \begin{tabular}{lccllllll}
    \cline{1-8}\cline{1-8}
    Detector  & $n$ & $F$       & Configuration & $f_{\rm lower}\,[{\rm Hz}]$ & $f_{\rm upper}\,[{\rm Hz}]$ & \reviewer{ Schedule}& \reviewer{Reference   }                                 &  \\\cline{1-8}
    DO-Conservative  & 2 &$\sqrt{3}/2$ & Triangle      & 0.001        & 10           &  \reviewer{2035-2050}     & \reviewer{Ref.~\cite{Sedda:2019uro, Sedda:2021yhn}                       }                    &  \\
    B-DECIGO         & 2 &$\sqrt{3}/2$ & Triangle      & 0.01               & 100   &\reviewer{2020s}              &\reviewer{Ref.~\cite{Yagi:2011wg}  }                                           &  \\
    AdvLIGO          & 2 & 1           & Rightangle    & 5                   & 5000   &\reviewer{O4   }          & \reviewer{LIGO documents\tnote{a}}   &  \\
    Voyager          & 2 & 1           & Rightangle    & 5                   & 10000      &\reviewer{Late 2020s }         &\reviewer{LIGO documents\tnote{b}  }&  \\
    CE-2             & 2 & 1           & Rightangle    & 3                   & 10000    &\reviewer{2040s  }         & \reviewer{Ref.~\cite{Reitze:2019iox,Reitze:2019dyk}    }                                     &  \\
    ET-D             & 3 &$\sqrt{3}/2$ & Triangle      & 1                   & 10000     &\reviewer{Mid 2030s }         & \reviewer{Ref.~\cite{Hild:2010id, Maggiore:2019uih}   }                                          &  \\
    \cline{1-8}
    \end{tabular}
    \begin{tablenotes}
      \item[a]{\small \url{ https://dcc.ligo.org/LIGO-T1800044/public}}
      \item[b]{\small \url{ https://dcc.ligo.org/LIGO-T1500293/public}}
    \end{tablenotes}
    \end{threeparttable}
\end{table}
%---------------------------------------------------------------------

\begin{table}
    \centering
    \renewcommand{\arraystretch}{1.25}
\caption{\reviewer{Number of observable BNSs and BBHs per year for different GW
detectors. The local BNS and BBH merger rates are chosen as $250\, {\rm
Gpc}^{-3} {\rm yr}^{-1}$ and $22\, {\rm Gpc}^{-3} {\rm yr}^{-1}$ respectively
~\cite{LIGOScientific:2021psn}. We adopt two models of merger rate's time
evolution, BNS/A~\cite{Schneider:2000sg,Cutler:2005qq} and
BNS/B~\cite{Madau:2014bja}, for BNSs, and one model for
BBHs~\cite{Baibhav:2019gxm}.}}\label{table2:}
    \begin{threeparttable}
    \begin{tabular}{lcccccc}
        \cline{1-7}\cline{1-7}
          & DO-Conservative & B-DECIGO & AdvLIGO & Voyager & CE-2   & ET-D   \\ \hline
    BNS/A & 12              & 384      & 50      & 5424    & 291264 & 161273 \\
    BNS/B & 13              & 421      & 52      & 6831    & 995039 & 394203 \\
    BBH & 70412           & 85792    & 866     & 34997   & 87021  & 87244 \\
    \cline{1-7}
    \end{tabular}
    \end{threeparttable}
\end{table}

In~\cref{fig:characteristic strain}, we illustrate the dimensionless
pattern-averaged characteristic strain $\tilde{h}_{c}(f)$ of GW150914-like and
GW170817-like signals, in contrast to the dimensionless noise spectral density
$\tilde{h}_{n}(f)$ of the above detectors \cite{Moore:2014lga}. The
``GW150914-like'' (``GW170817-like'') signal indicates mock GW event that adopts
the component masses and distance of real GW150914 (GW170817), but omits the
spin effects and uses the angle-averaged waveform in
\cref{eqn:amplitude,eqn:phase_GR}. The quantities $\tilde{h}_c(f)$ and
$\tilde{h}_n(f)$ are defined by $2 f \big|\tilde{h}(f)\big|$ and $\sqrt{f
S_{n}^{\rm eff}(f)}$ respectively, where $S_{n}^{\rm eff}$ is the effective
noise PSD, $S_{n}^{\rm eff}(f)= {S_{n}(f)} / {n F^2}$, where $n$ and $F$ are the
equivalent numbers and the form factor of the detectors. Configurations of
detectors are collected in~\cref{tab:curves}. In the figure  SNRs are
proportional to the area between the signals and the noise
curves~\cite{Moore:2014lga}.

\begin{figure}%[htp]
    \centering\hspace{0cm}
    \begin{subfigure}[c]{0.48\textwidth}
        \includegraphics[width=7.4cm]{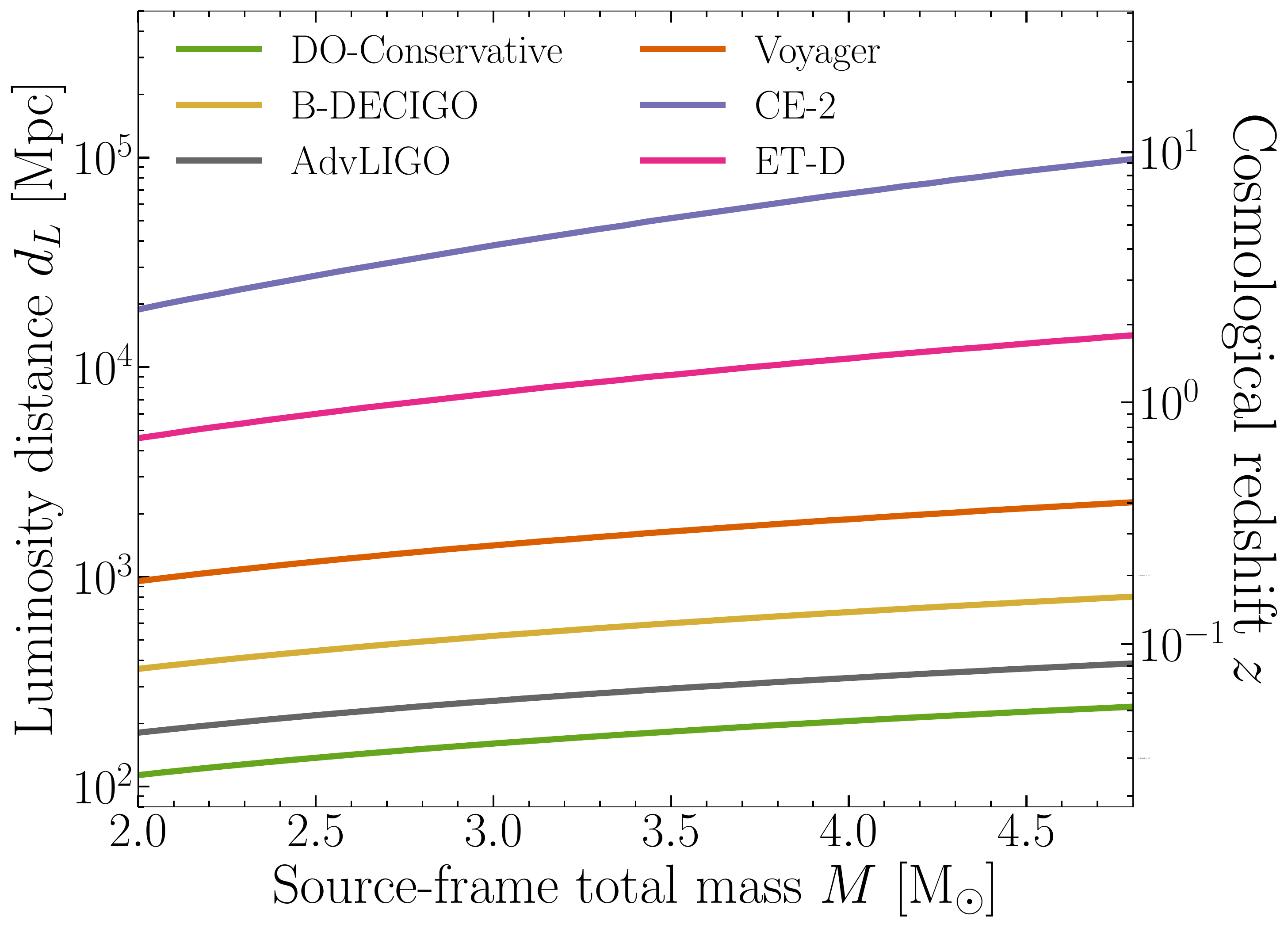}
    \end{subfigure}\hspace{-0.1cm}
    \begin{subfigure}[c]{0.48\textwidth}
        \includegraphics[width=7.4cm]{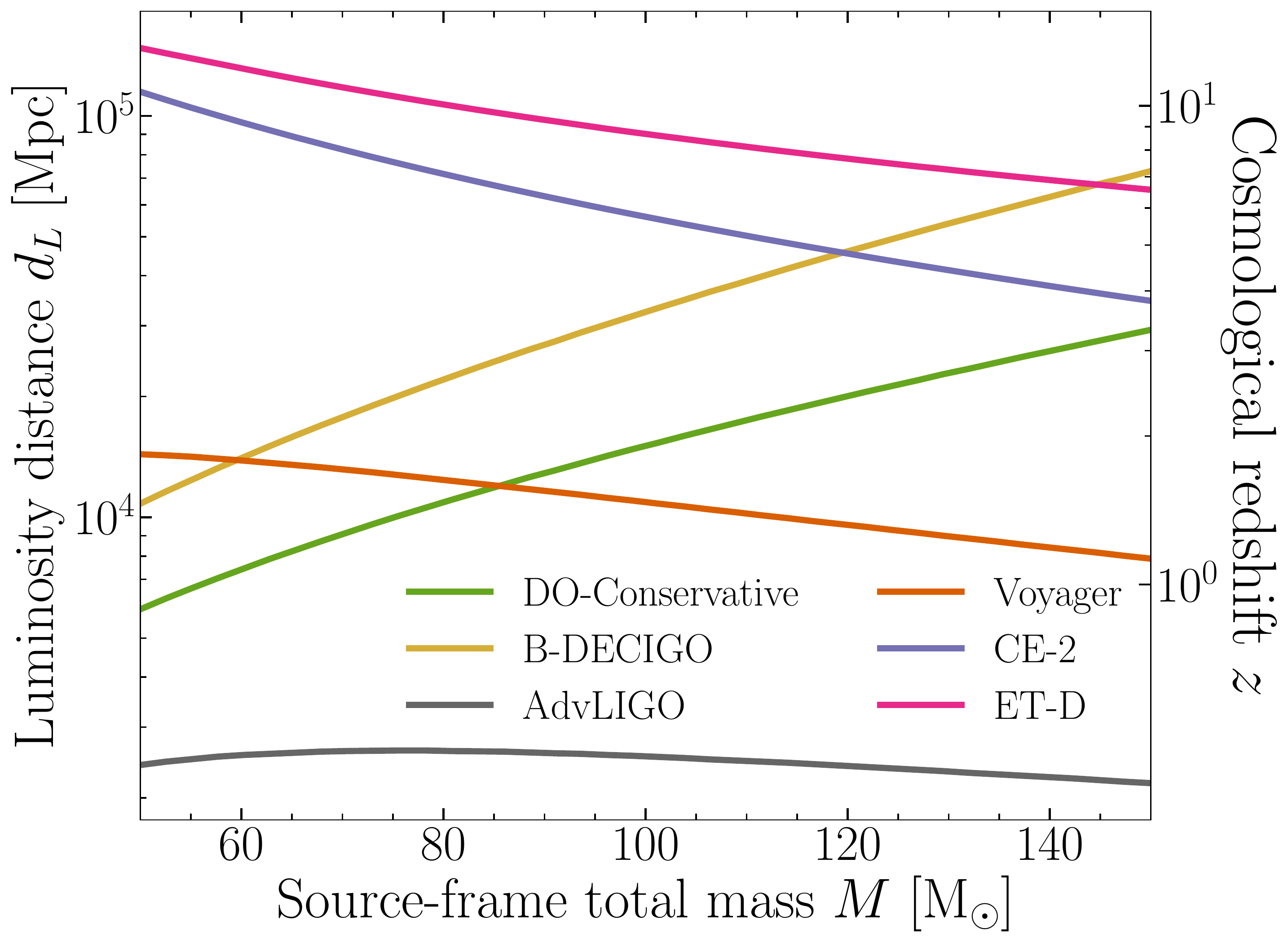}
    \end{subfigure}
    \caption{\reviewer{ Cosmological reach to equal-mass BNSs ({\it left}) and
    BBHs ({\it right}) for different detectors. The $\Lambda$CDM parameters used
    for calculation are matter density $\Omega_{\rm m} \approx 0.315$, dark
    energy density $\Omega_\Lambda \approx 0.685$, and the Hubble constant
    $H_0\approx 67.4\ \text{km}\,\text{s}^{-1}\,\text{Mpc}^{-1}$
    \cite{Planck:2018vyg}. The thresholds of detection are set to $\rho =8$ and
    $\rho=15$ for ground-based and space-based detectors, respectively. }}
    \label{fig:horizon}
\end{figure}

\reviewer{ The cosmological reach is defined as the corresponding luminosity
distance to the detection threshold. In \cref{fig:horizon}, we plot cosmological
reach to equal-mass BNSs and BBHs for different detectors. With the cosmological
reaches, we estimate the corresponding BNS and BBH event rates in
\cref{table2:}, showing the prospect that a large population of BNSs and BBHs
can be detected by the above detectors. }

%---------------------------------------------------------------------
\section{Simultaneous bounds on $B$ and $\dot G$}
\label{sec:bounds}
%---------------------------------------------------------------------

%improvement
We will use FIM to obtain simultaneous constraints on DGR and varying-$G$
effects. Let indices $i$ and $j$ in the FIM run through the set of parameters
$\{\ln \mathcal{A}, \ln \eta, \ln \mathcal{M}, t_{c}, \Phi_{c}, B, \dot G\}$. We
calculate partial derivatives analytically from the GW
waveform~\eqref{eqn:waveform}, and  they are shown in~\ref{sec:appendix}.
Substituting values of source parameters together with fiducial values
$B=0$ and $\dot G=0$, we calculate the FIM through \cref{eqn:fisher}.  Then, the
standard deviations of $B$ and $\dot{G}$ are obtained from corresponding
elements of the inverse of FIM, providing  1-$\sigma$ bounds on DGR and
varying-$G$ effects. To convert between luminosity distance and redshift, we
adopt the $\Lambda$CDM model with Hubble constant $H_0\approx 67.4\
\text{km}\,\text{s}^{-1}\,\text{Mpc}^{-1}$,  matter density fraction
$\Omega_{\rm m} \approx 0.315$, and  vacuum energy density fraction
$\Omega_\Lambda \approx 0.685$ \cite{Planck:2018vyg}. 

Considering how the waveform depends on different parameters before calculation
can bring some simplification. \reviewer{The luminosity distance $d_L$ does not appear in
the phase but in the amplitude as shown in~\cref{eqn:amplitude}.  According to Eqs. (\ref{eqn:fisher}) and (\ref{eqn:FIMsigma}), we approximately have $F_{ij} \propto 1/d^2_L$ and $\sigma_B$ (or $ \sigma_{\dot{G}}$) $\propto
d_L$.} For BNS events, we use $d_L = 40\, {\rm Mpc}$, while for BBH events we use
$d_L = 440\, {\rm Mpc}$. In addition, according to \cref{eqn:phase_GR}, different
values of $t_c$ and $\Phi_c$ do not affect the constraints, and we use fiducial
values $t_c=0$ and $\Phi_c=0$.

%---------------------------------------------------------------------
\begin{figure}%[htp]
    \centering 
    \includegraphics[width=7cm]{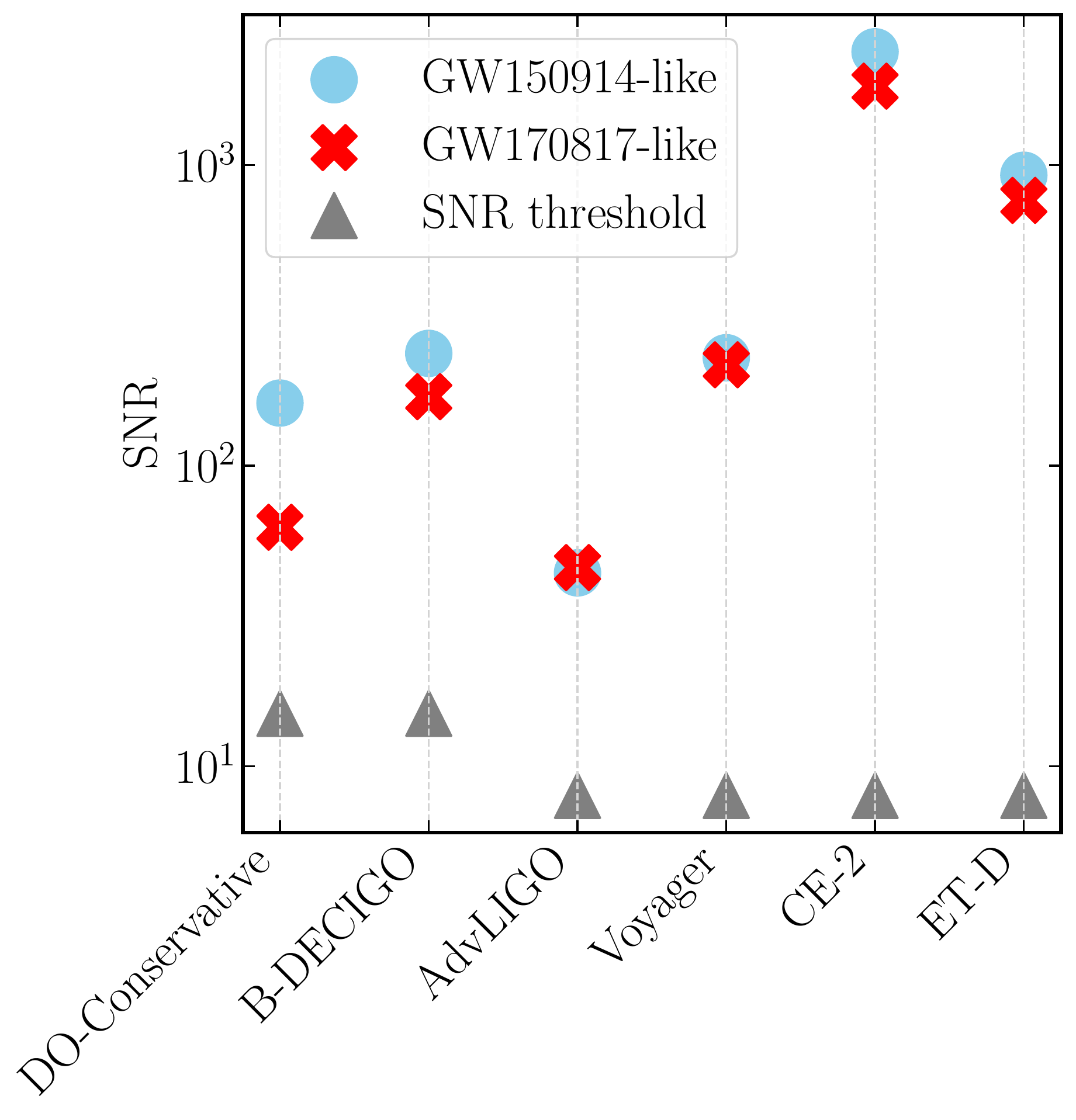} 
    \caption{SNRs of GW150914-like and GW170817-like events in different
    detectors. The gray solid triangles stand for the thresholds of detection
    with $\rho =8$ and $\rho=15$ for ground-based and space-based detectors
    respectively. }
    \label{fig:curves_pulsar_SNR}
\end{figure}
%---------------------------------------------------------------------

%---------------------------------------------------------------------
\begin{figure}%[htp]
    \centering
    \begin{subfigure}[c]{0.48\textwidth}
        \includegraphics[width=7cm]{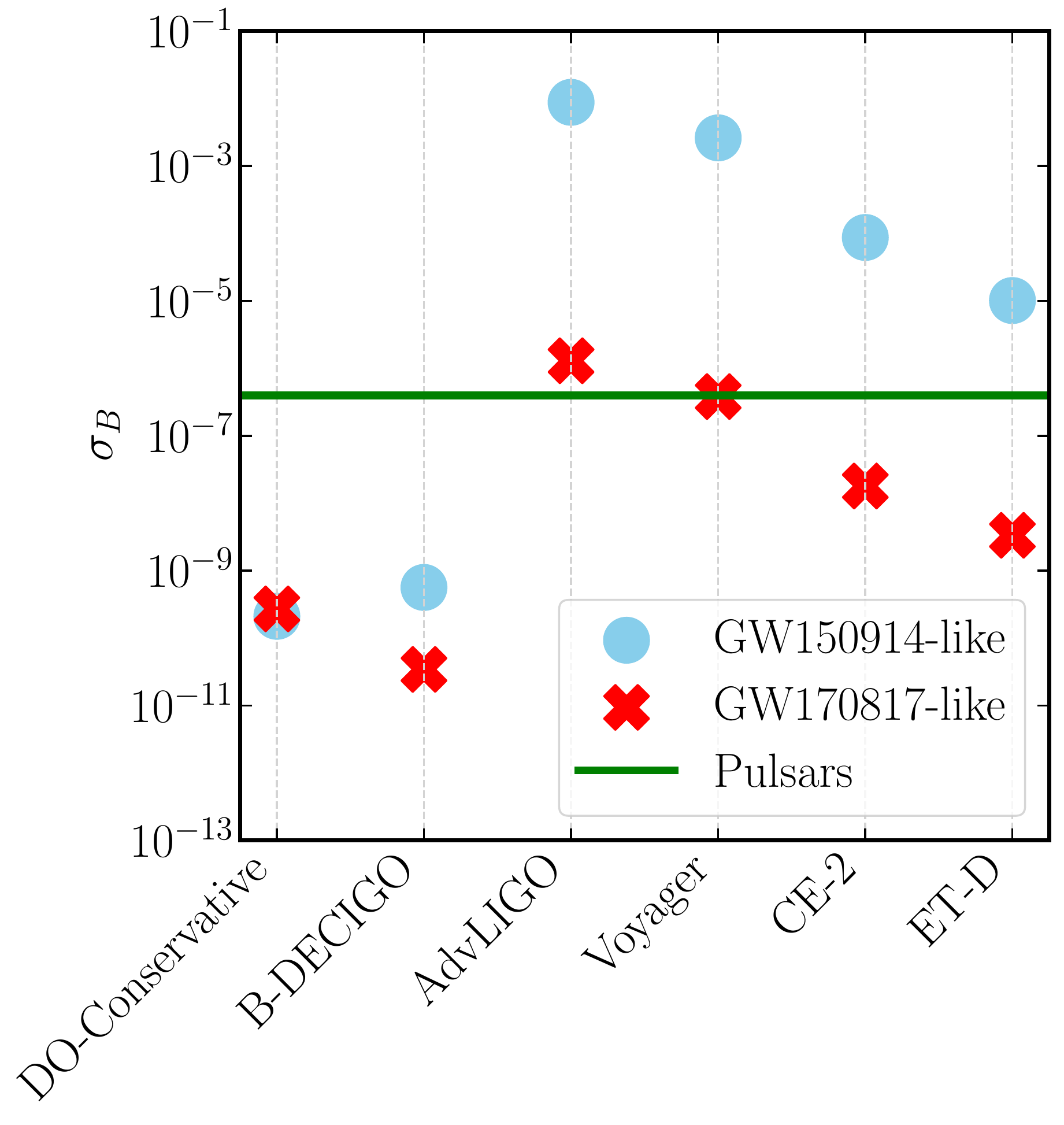}
        \label{fig:curves_pulsar_B}
    \end{subfigure}\hspace{1em}%
    \begin{subfigure}[c]{0.48\textwidth}
        \includegraphics[width=7cm]{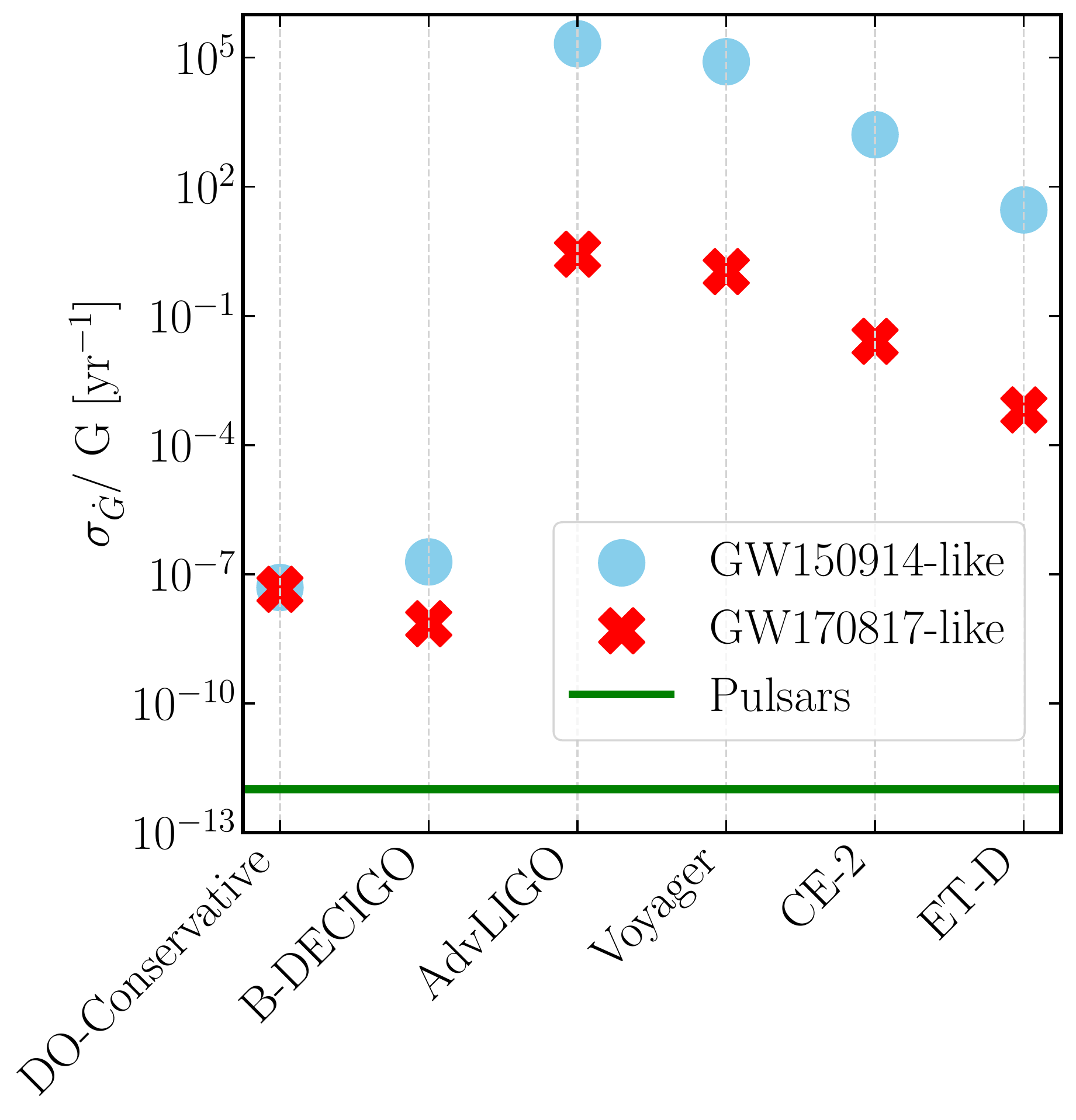}
        \label{fig:curves_pulsar_G}
    \end{subfigure}
    \caption{A comparison of $\sigma_B$ and $\sigma_{ \Dot G}/G$ in detections
    of GW150914-like and GW170817-like events using different detectors.
    Current timing results of pulsars are shown with green lines for comparison. }
    \label{fig:curves_pulsar}
\end{figure}
%---------------------------------------------------------------------

%---------------------------------------------------------------------
\subsection{Constraints on $B$ and $\dot G$}
%---------------------------------------------------------------------

First, we investigate the constraints on $B$ and $\dot G$ from different detectors.
\Cref{fig:curves_pulsar_SNR} plots SNRs of GW150914-like and GW170817-like
events in different detectors~\cite{Moore:2019pke}. The SNRs in all detectors are so
high that we can use the FIM to estimate the measurement precision of parameters
to a good extent~\cite{Vallisneri:2007ev}.

\Cref{fig:curves_pulsar} shows the constraints on $B$ and $\dot{G}$ with our two
example events in different detectors. They are compared with the bounds from
binary pulsars~\cite{Zhu:2018etc}.  The space-based GW detectors DO-Conservative
and B-DECIGO outperform the other GW detectors even though they do not have the
highest SNRs. This can be explained by the lower accessible frequency of these
detectors and contributes more to the negative $-1$\,PN and $-4$\,PN terms in
$\Psi (f)$. 
For the GW170817-like event, the current GW detector AdvLIGO cannot
constrain $B$ as tightly as pulsars, while future detectors like
DO-Conservative, B-DECIGO, CE, and ET will provide bounds 2 to 4 orders of
magnitude tighter. As for $\dot{G}$, the bounds by GW detectors from the
GW170817-like event are looser than $\sigma_{ \Dot G}/G < 10^{-12} \, {\rm
yr}^{-1}$ (68\% CL) from pulsars~\cite{Zhu:2018etc}. It is caused by the even
more negative PN order of the varying-$G$ effect in the phase correction.
Constraints from the GW150914-like event are close to those from the
GW170817-like event for space-based GW detectors, but much looser for
ground-based ones. Comparing to space-based detectors,  there are greater
differences of the GW cycles in the frequency bands between BBH and BNS
inspirals for ground-based detectors.\footnote{More strictly, this attributes to
the differences in the integral intervals, $\left[f_{\min}, f_{\max} \right]$,
between two kinds of detectors (see \cref{fig:characteristic strain}).} As
mentioned in Sec.~\ref{sec:intro}, though limits from BBHs are not as tight as
those from BNSs, they are still meaningful, because observations of pulsars
cannot give any limits for BBHs at all \cite{Barausse:2016eii}. 

\begin{figure}%[htp]
    \centering\hspace{-2.5cm}
    \begin{subfigure}[c]{0.48\textwidth}
        \includegraphics[width=10cm]{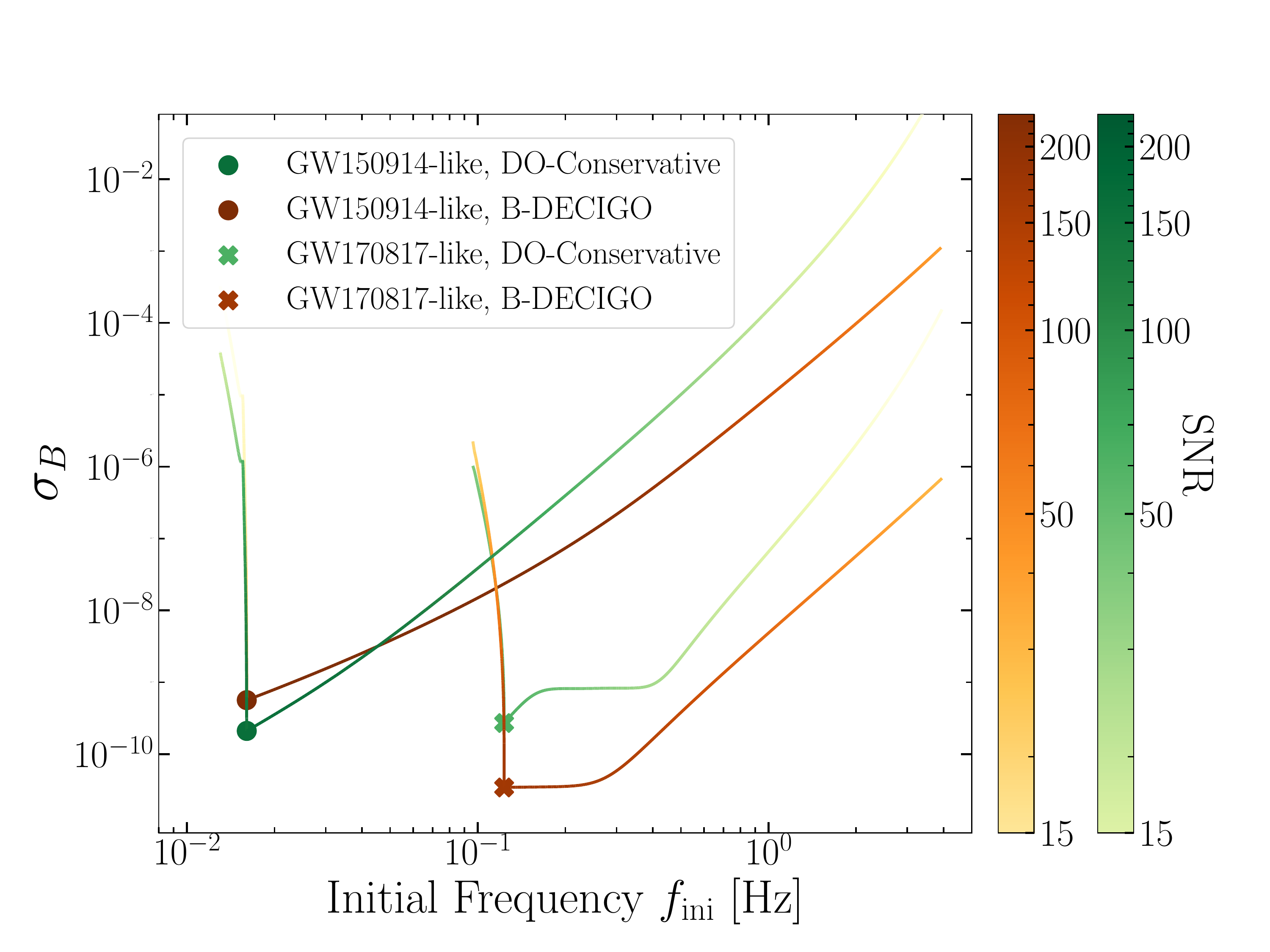}
    \end{subfigure}\hspace{2cm}
    \begin{subfigure}[c]{0.48\textwidth}
        \includegraphics[width=10cm]{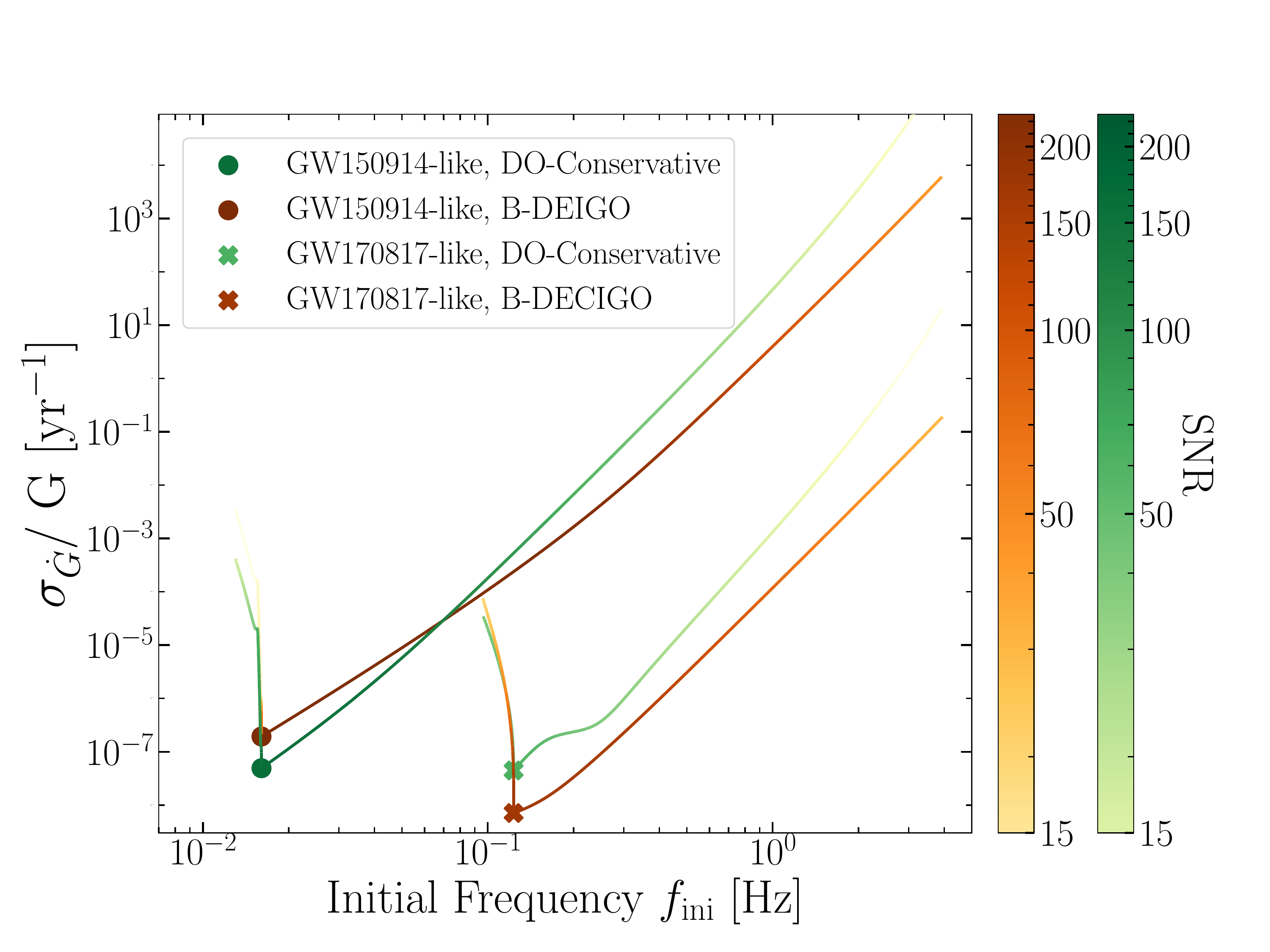}
    \end{subfigure}
    \caption{\reviewer{ Dependence of $\sigma_B$ ({\it left}) and
    $\sigma_{\dot{G}}$ ({\it right}) on the initial frequency in DO-Conservative
    and B-DECIGO detectors, for GW150914-like and GW170817-like events. The
    color of lines represents the value of the SNR. The $f_{\rm 4\,yr}$ of
    GW150914-like and GW170817-like events, as well as the corresponding
    constraints, are marked with ``$\bullet$'' and ``$\times$''.}}
    \label{fig:ini_f}
\end{figure}

\reviewer{ When calculating the SNRs in \cref{fig:curves_pulsar_SNR} and the
constraints in \cref{fig:curves_pulsar}, we have chosen $f_{\max} = \min
\left\{f_{\rm upper}, f_{\rm ISCO}\right\}$ and $f_{\min} = \max \left\{f_{\rm
lower}, f_{\rm 4\,yr}\right\}$ for space-based detectors, which correspond to
the best constraints potentially achievable by those future detectors for the
two GW events. In fact, an observed binary will have an initial frequency
according to the underlying population, and any other initial frequency
different from $f_{\rm min}$ will lead to a lower SNR and looser
constraints~\cite{Barbieri:2022zge}. In \cref{fig:ini_f}, we plot how the SNR
and constraints depend on the initial frequencies of GW150914-like and
GW170817-like events for space-like detectors DO-Conservative and B-DECIGO. When
the initial frequency $f_{\rm ini}$ is larger than $f_{\rm min}$, the integral
range will be $[f_{\rm ini}\,,f_{\max}]\subsetneqq [f_{\rm 4\,yr}\,,f_{\max}]$,
corresponding to a lower SNR and looser limits. For those $f_{\rm ini}$ smaller
than $f_{\rm min}$, though the DGR and varying-$G$ effects are more significant
for lower frequencies, it also takes more time for each circle of the binary in
orbital evolution. Limited by the assumed 4-year lifetime of the space-based
detectors, for $f_{\rm ini}<f_{\rm 4\,yr}$ we still have lower SNRs and looser
constraints, as shown in \cref{fig:ini_f}. Moreover, for those frequencies far
away from $f_{\rm 4\,yr}$, the SNRs become even smaller than the detection
threshold, not to mention the FM approximation and constraints. }

%---------------------------------------------------------------------
%---------------------------------------------------------------------

%引入
Then we investigate how a binary's parameters influence $\sigma_B$ and
$\sigma_{\dot{G}}$.  We choose ET-D and B-DECIGO as the representatives of
ground-based and space-based detectors, respectively. We expect the properties
of parameter dependence to be alike within the same kind of detector due to
their similar sensitivity ranges and shapes.

% \subsection{$\sigma_{ B}$} 
%引入

\begin{figure}[h]
    \centering
    \begin{subfigure}[c]{0.45\textwidth}
        \includegraphics[width=0.8\textwidth]{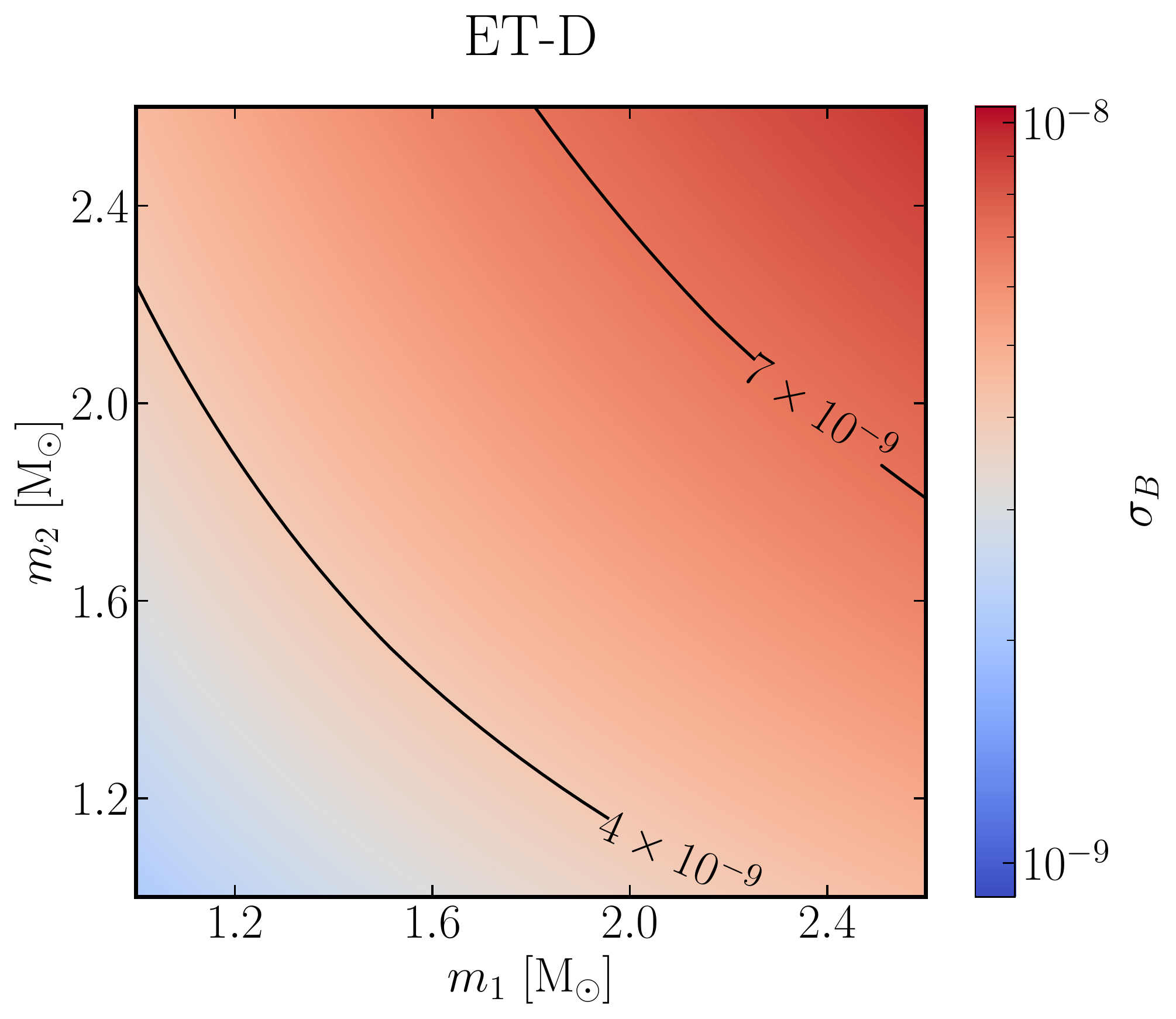}
        \label{fig:ET-D_sigmaB_BNS}
    \end{subfigure}\hspace{1em}%
    \begin{subfigure}[c]{0.45\textwidth}
        \includegraphics[width=0.8\textwidth]{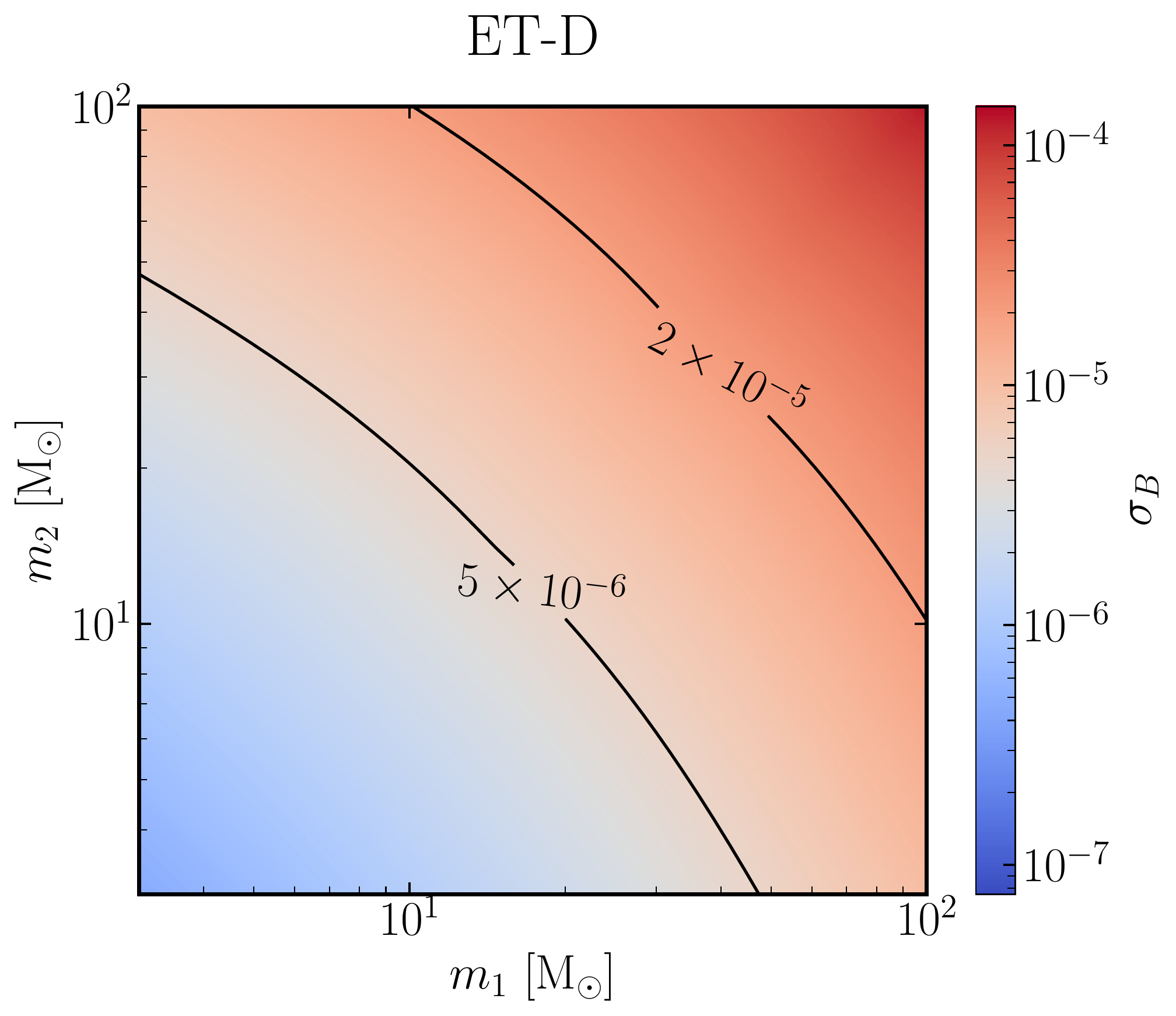}
        \label{fig:ET-D_sigmaB_BBH}
    \end{subfigure}\hspace{1em}%
    \\
    \begin{subfigure}[c]{0.45\textwidth}
        \includegraphics[width=0.8\textwidth]{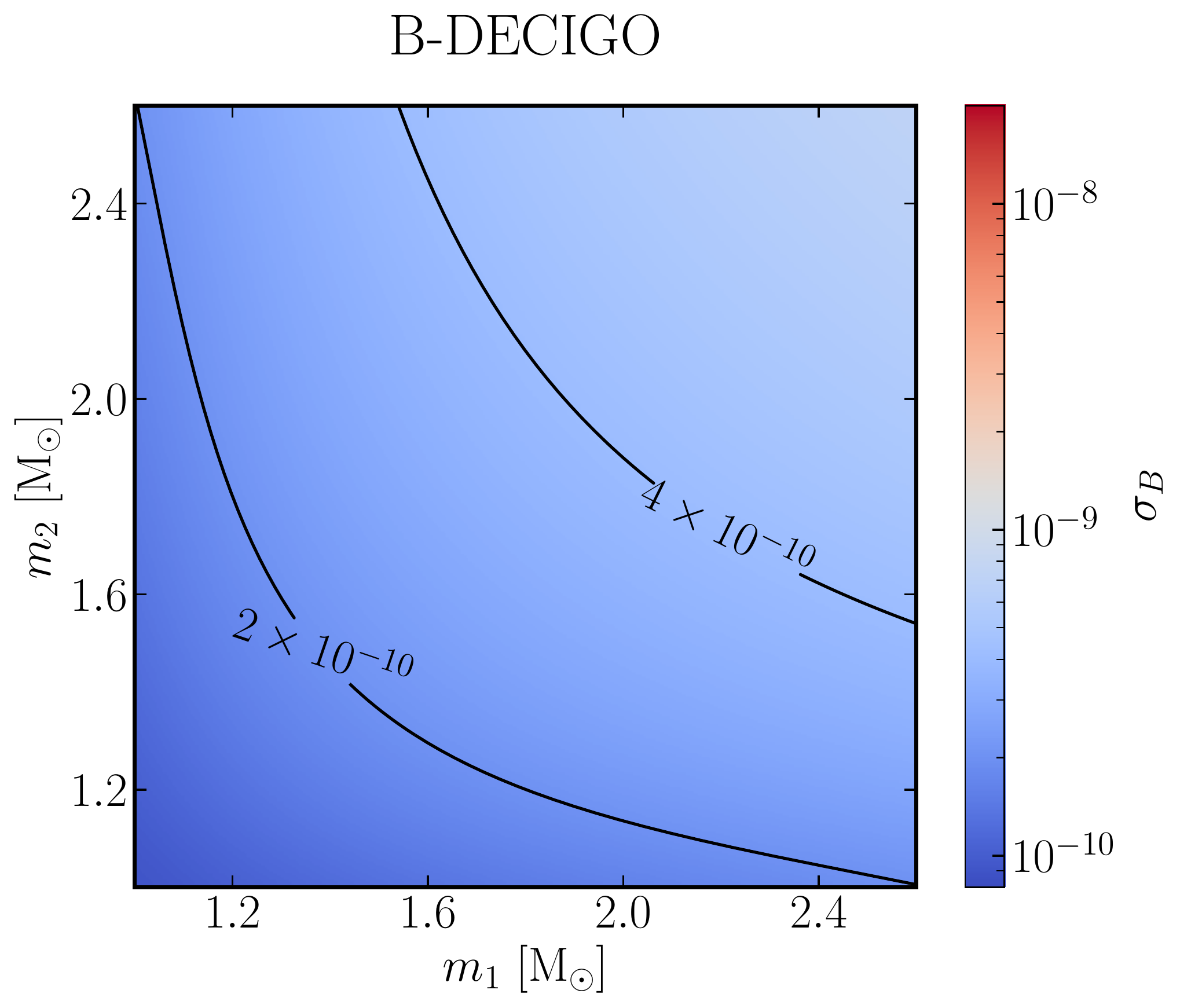}
        \label{fig:B-DECIGO_sigmaB_BNS}
    \end{subfigure}\hspace{1em}%
    \begin{subfigure}[c]{0.45\textwidth}
        \includegraphics[width=0.8\textwidth]{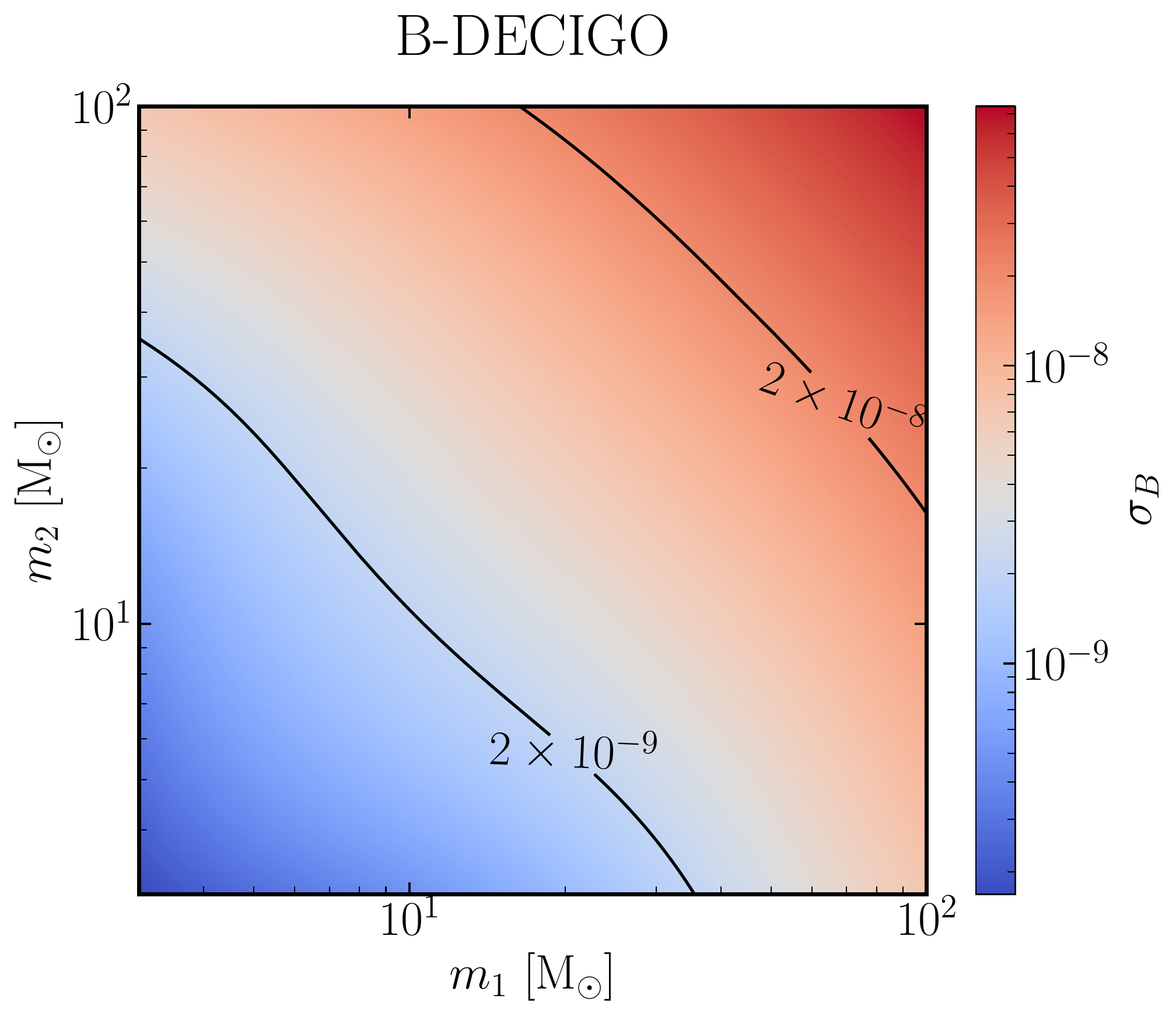}
        \label{fig:B-DECIGO_sigmaB_BBH}
    \end{subfigure}
    \caption{Dependence of $\sigma_B$ on component masses in two GW detectors,
    ET-D ({\it upper}) and B-DECIGO ({\it lower}), for BNSs ({\it left}) and BBHs ({\it right}).
    } \label{fig:sigmaB}
\end{figure}

The constraints on $B$ are shown in~\cref{fig:sigmaB}. Here, we vary the
component masses $m_1$ and $m_2$.  Systems with smaller masses give slightly
tighter  bounds on $B$. A lighter system has a larger separation when it enters
the detector's sensitivity band, resulting in slower damping and more cycles
within the detector bandwidth, and a tighter limit in the end. In
\cref{fig:sigmaB}, the limits on $B$ only weakly rely on the components of BNSs,
varying within one order of magnitude, and all limits are tighter than the
pulsar results~\cite{Zhu:2018etc}. For BBHs, there exists strong dependence
between $\sigma_B$ and BBHs' component masses. Note that in \cref{eqn:phase_-1}
the $-1$\,PN phase term does not explicitly depend on mass parameters, and the
dependence of mass actually originates from the range of integral. A GW signal
from a massive BBH will stay in the detector for fewer cycles, leading to looser
constraints. This dependence is particularly evident in ET-D which has a
relatively high frequency band.

%---------------------------------------------------------------------
%---------------------------------------------------------------------

%---------------------------------------------------------------------
\begin{figure}[htp]
    \centering
    \begin{subfigure}[c]{0.45\textwidth}
        \includegraphics[width=0.8\textwidth]{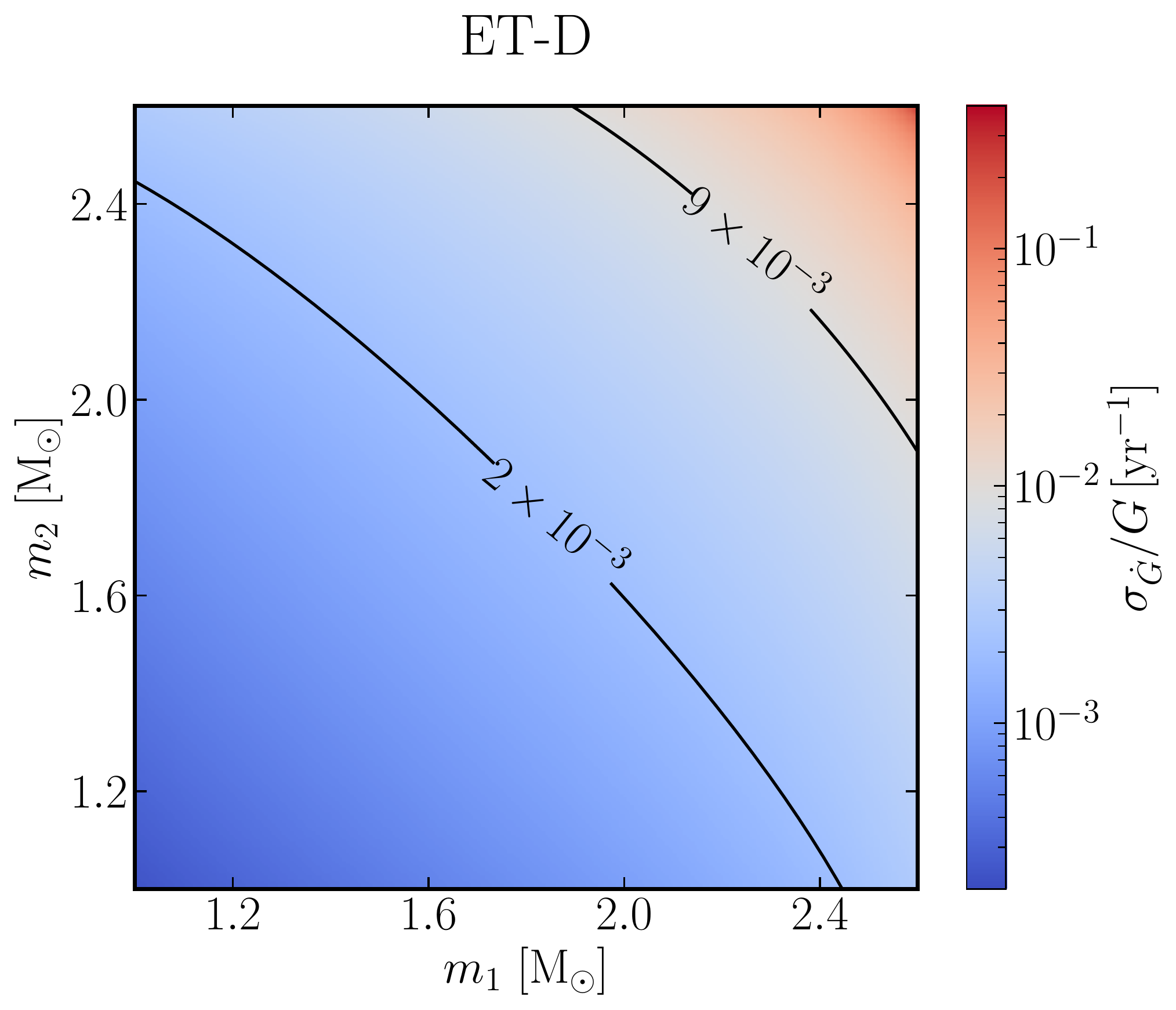}
        \label{fig:ET-D_sigmaGdot_BNS}
    \end{subfigure}\hspace{1em}%
    \begin{subfigure}[c]{0.45\textwidth}
        \includegraphics[width=0.8\textwidth]{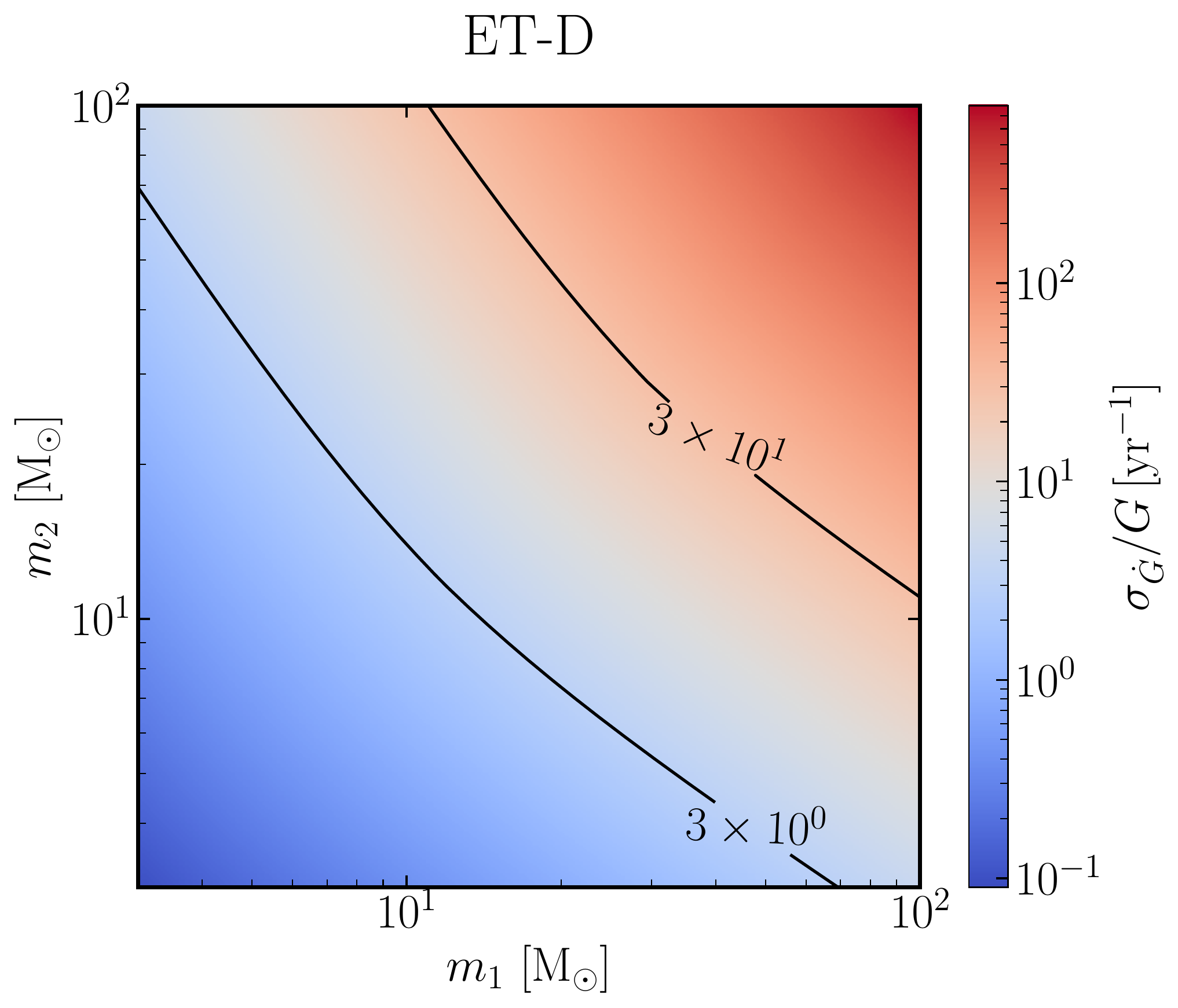}
        \label{fig:ET-D_sigmaGdot_BBH}
    \end{subfigure}\hspace{1em}%
    \\
    \begin{subfigure}[c]{0.45\textwidth}
        \includegraphics[width=0.8\textwidth]{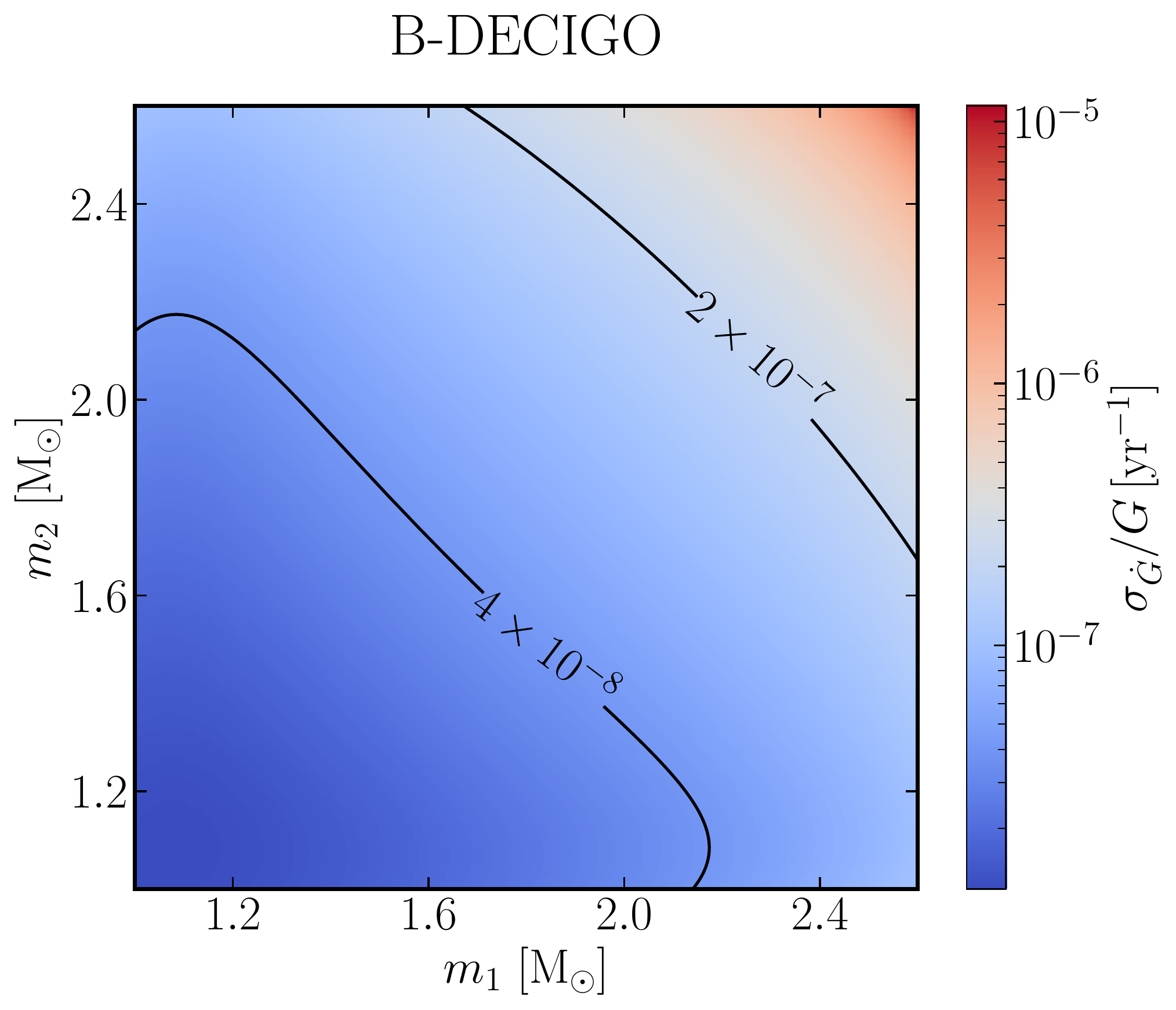}
        \label{fig:B-DECIGO_sigmaGdot_BNS}
    \end{subfigure}\hspace{1em}%
    \begin{subfigure}[c]{0.45\textwidth}
        \includegraphics[width=0.8\textwidth]{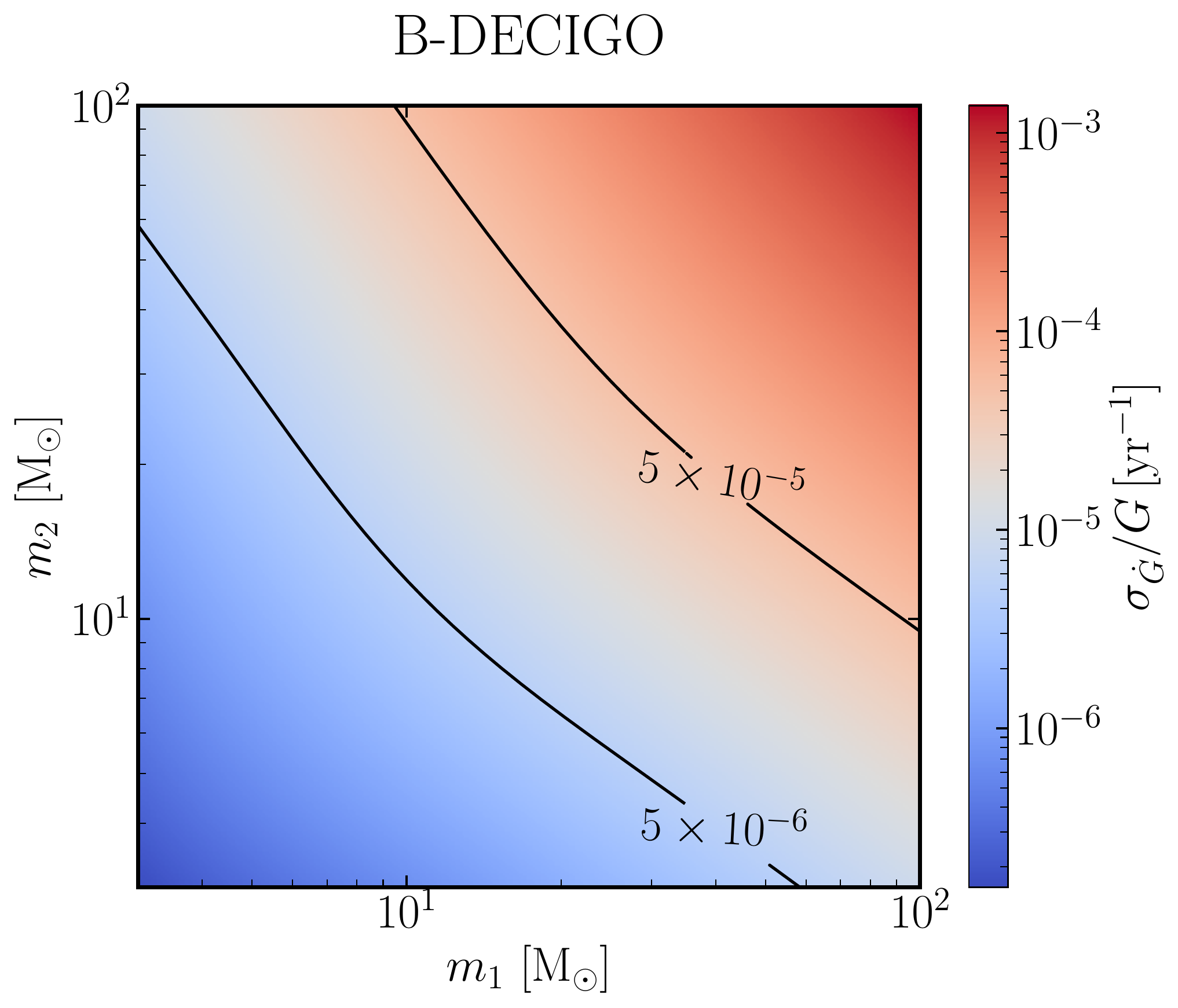}
        \label{fig:B-DECIGO_sigmaGdot_BBH}
    \end{subfigure}
    \caption{Same as \cref{fig:sigmaB}, for $\sigma_{ \Dot G}$.}
    \label{fig:sigmaGdot}
\end{figure}
%---------------------------------------------------------------------

\begin{figure}
    \centering
    \begin{subfigure}[c]{0.49\textwidth}
        \includegraphics[width=8cm]{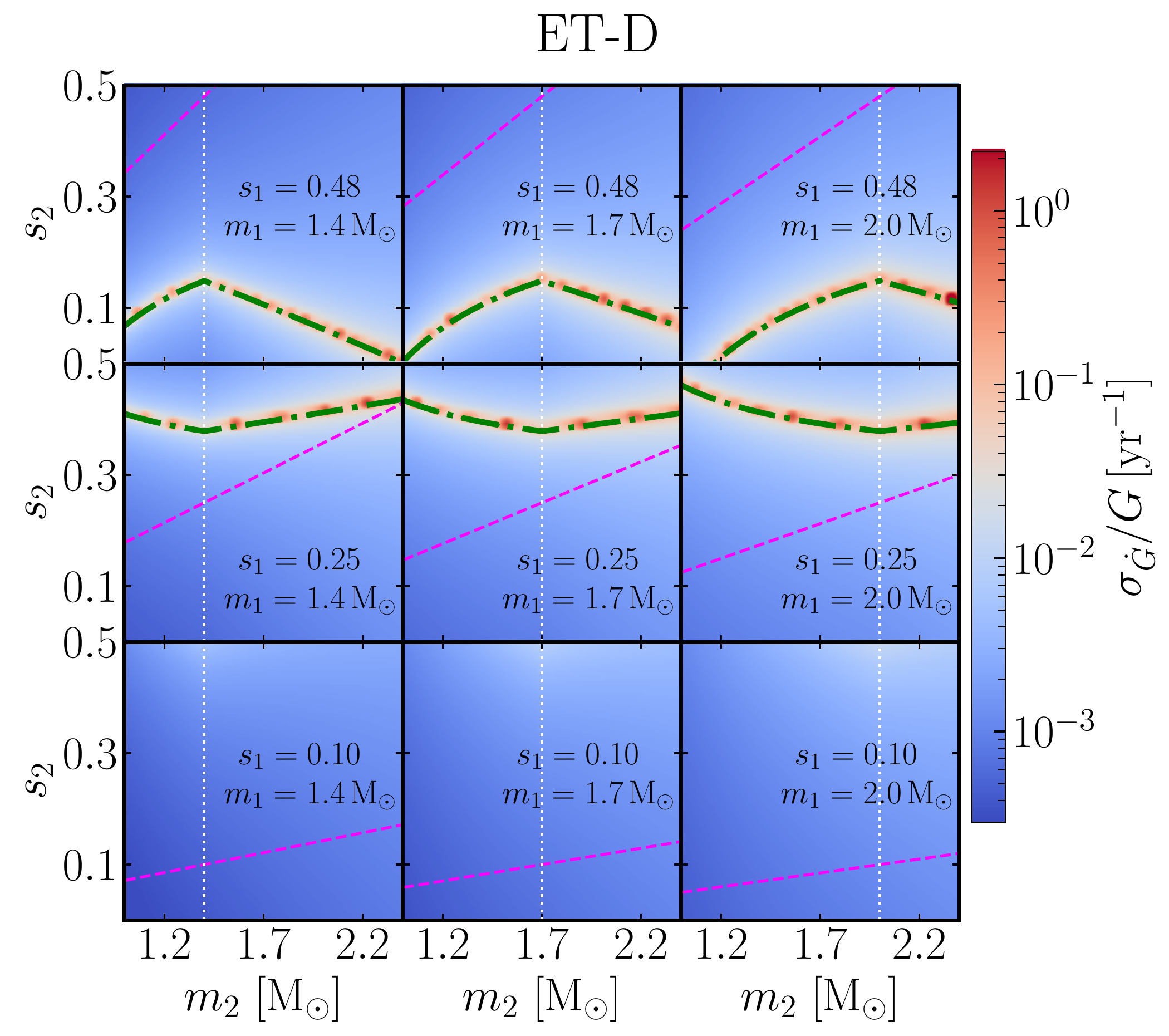}
        \label{fig:ET-D_sigmadotG_m1m2}
    \end{subfigure}
    \begin{subfigure}[c]{0.49\textwidth}
        \includegraphics[width=8cm]{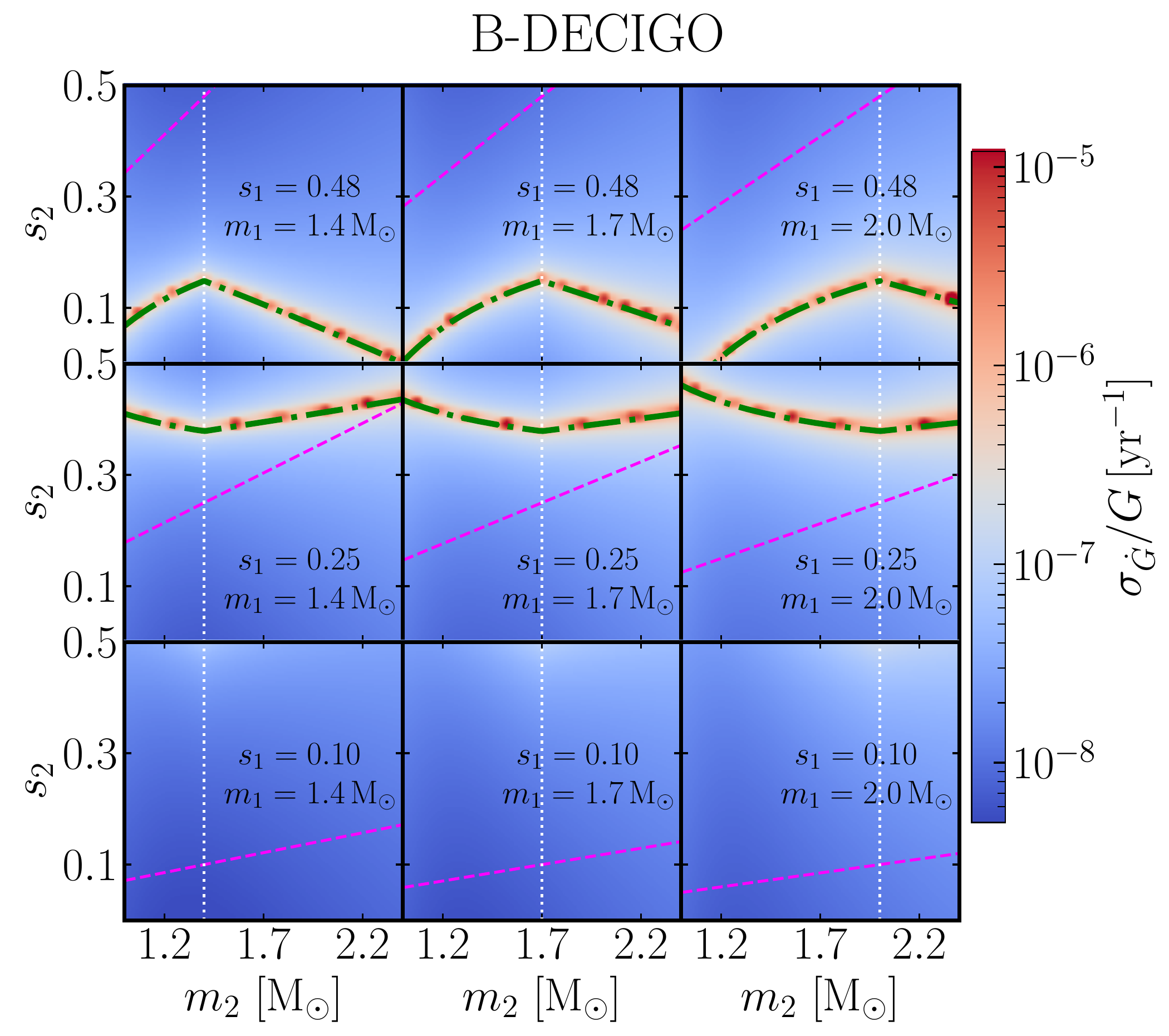}
        \label{fig:B-DECIGO_sigmadotG_m1m2}
    \end{subfigure}
    \caption{Dependence of $\sigma_{ \Dot G}$ on component masses and
    sensitivities of BNSs. The green dot-dashed lines correspond to 
    $\mathfrak{F}=0$ [see Eq.~(\ref{eq:frakF})], where FIM becomes singular. The
    fuchsia dashed lines are plotted with the assumption of $s_2 = m_2 \cdot
    s_1/m_1.$ The white dotted lines show the primary masses of binaries.}
    \label{fig:sigmadotG}
\end{figure}

The phase correction term  of varying-$G$ in~\cref{eqn:phase_-4} is relatively
complicated, dependent on four parameters, $m_1$, $m_2$, $s_1$, and $s_2$.
In~\cref{fig:sigmaGdot}, we vary parameters $m_1$ and $m_2$ in BNS and BBH
systems like \cref{fig:sigmaB}, and adopt the approximation in
\cref{eqn:sensitivity} for NSs and fix the sensitivity as $s=0.5$ for BHs. To
study the parameter space outside \cref{eqn:sensitivity}, as discovered in some
alternative gravity theories, in \cref{fig:sigmadotG} we release the
approximation in \cref{eqn:sensitivity} and vary all the four parameters for BNS
systems. Same as constraining $B$, binaries with larger masses generally give a
larger $\sigma_{\dot{G}}$. From our results, all limits on $\sigma_{\dot{G}}$
are looser than results from pulsars~\cite{Zhu:2018etc}, because the GW
detectors here can hardly reach the low frequency where ${\dot{G}}$'s $-4$\,PN
effects are as significant as in binary pulsars.

In \cref{fig:sigmadotG}, $\sigma_{ \Dot G}$ turns out to be very large in some
striped-like regions. This is due to singular FIMs. In these regions, $ \partial
\tilde{h}/{\partial {\dot G}} $ almost vanishes. Therefore, FIM is nearly
singular, and the corresponding element in the covariance matrix becomes
extremely large. To show this, we define a factor ``$\mathfrak{F}$'' as 
%--
\begin{align}
    \label{eq:frakF}
    \mathfrak{F}\equiv\frac{\mathcal{M}}{\eta^{13/5}} \left( 11 - \frac{35}{2} S
    - \frac{41}{2} \sqrt{1 - 4 \eta} \Delta S  \right)\propto \frac{\partial
    \tilde{h}(f)}{\partial {\dot G}}\,,
\end{align}
%--
and plot lines where $\mathfrak{F}= 0$ in~\cref{fig:sigmadotG}. As expected, the
regions with large $\sigma_{ \Dot G}$ and the lines where $\mathfrak{F}= 0$ are
highly consistent. $\mathfrak{F}$ only depends on properties of a binary, so
regions with large $\sigma_{ \Dot G}$ in different detectors have the same
shape, with differences only in amplitude. Singular (or quasi-singular) FIMs
cannot provide credible forecasts of parameter estimation, and alternative
analysis methods are required \cite{Vallisneri:2007ev, Wang:2022kia}. Though the
constraint on $\Dot G$ becomes extremely loose in the adjacency of the
singularities, in most area of the parameter space, $\sigma_{ \Dot G}$ remains
to be small. Therefore, BNSs are still in general good sources for constraining
$\sigma_{\Dot G}$, as long as $\mathfrak{F}$ is not close to zero.  For BBHs,
their sensitivities are fixed as 0.5, so their $\mathfrak{F}$ never equals to
zero.  Same as the constraints on $B$, BBH inspirals in our study cannot give
limits as tight as BNSs and pulsars.

%---------------------------------------------------------------------
%---------------------------------------------------------------------

%---------------------------------------------------------------------
\subsection{The role of correlation}
%---------------------------------------------------------------------
%%

Though $B$ and $\Dot G$ show up at different PN orders in the modified GW phase,
they are correlated in the posterior distributions. In FIM analysis, the
correlation between them is characterized by $c_{B\dot G}$
using~\cref{eqn:FIMsigma}.  The sign of $c_{B\dot G}$ changes when crossing the
singular region mentioned above, rooting in the sign changing of $\mathfrak{F}$.
When calculating the correlation coefficients, we adopt \cref{eqn:sensitivity}
for BNSs, and fix $s = 0.5$ for BBHs. For BBHs (whose masses range from
$3\,{M}_{\odot}$ to $100\,{M}_{\odot}$), $\big| c_{B\dot G} \big|$ is
almost always greater than 0.9, while for BNSs, $\big| c_{B\dot G} \big|$ is
more dependent on the specifics of detectors and system masses. Only in a small
region of the parameter space, $\big| c_{B\dot G} \big|$ is smaller than 0.5.
Generally, the correlation between $B$ and $\Dot G$ is smaller for BNS inspirals
than BBH inspirals, and space-based detectors give a smaller correlation than
ground-based detectors. For BNSs, the space-based detectors can distinguish
$-1$\,PN and $-4$\,PN effects better with a lower frequency band, giving a
relatively small correlation. But in a BBH inspiral, $\Dot G$'s $-4$\,PN effects
hardly enter the frequency bands of both ground-based and space-based detectors,
leading to a strong correlation between $B$ and $\Dot G$. In all, for almost any
signal from BNSs or BBHs, we expect $B$ and $\Dot G$ to be strongly correlated.

\begin{figure}[h]
    \centering
    \begin{subfigure}[c]{0.4\textwidth}
        \includegraphics[width=0.78\textwidth]{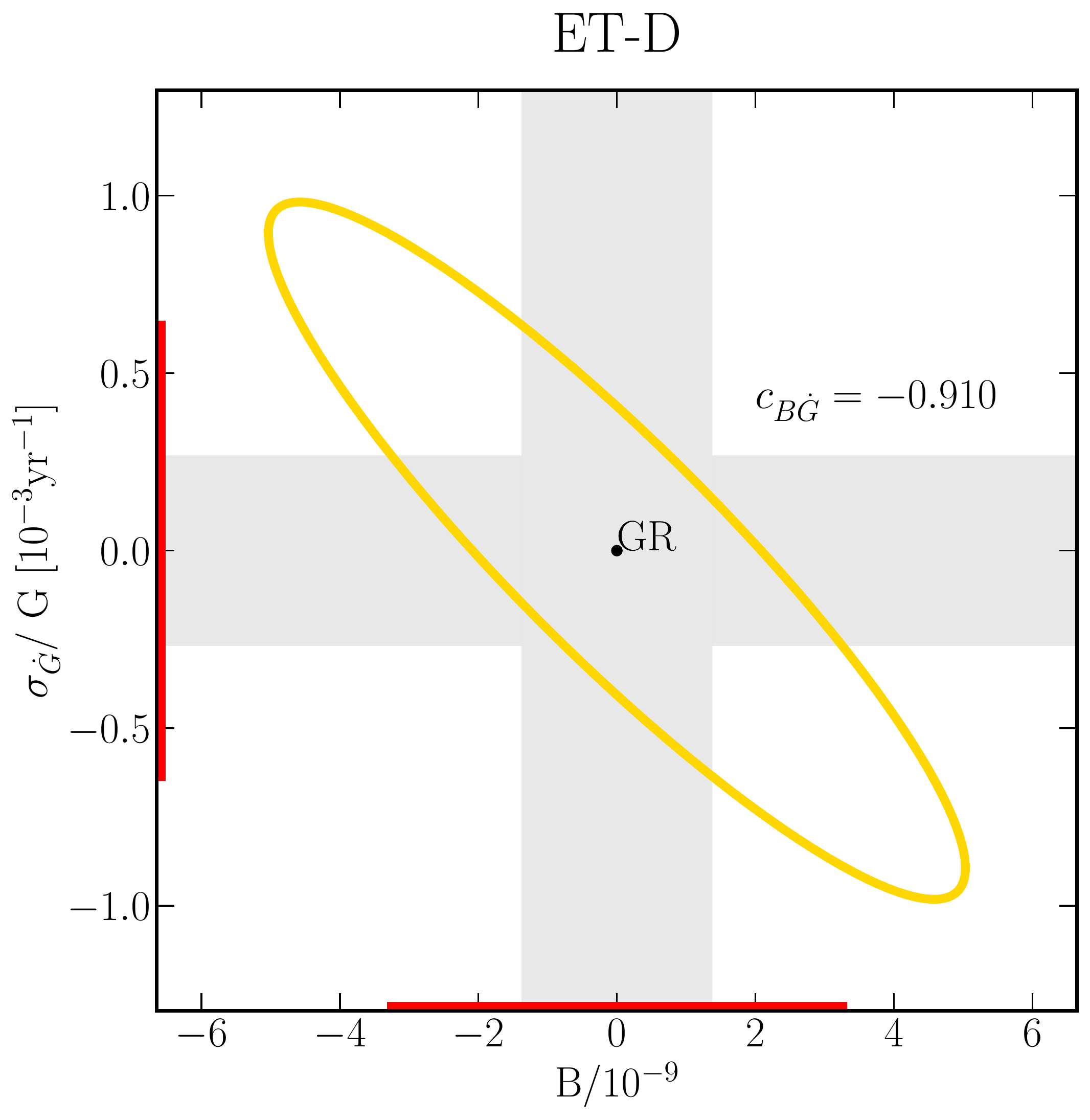}
        \label{fig:correlation_ET-D_BNS}
    \end{subfigure}\hspace{1em}%
    \centering
    \begin{subfigure}[c]{0.4\textwidth}
        \includegraphics[width=0.78\textwidth]{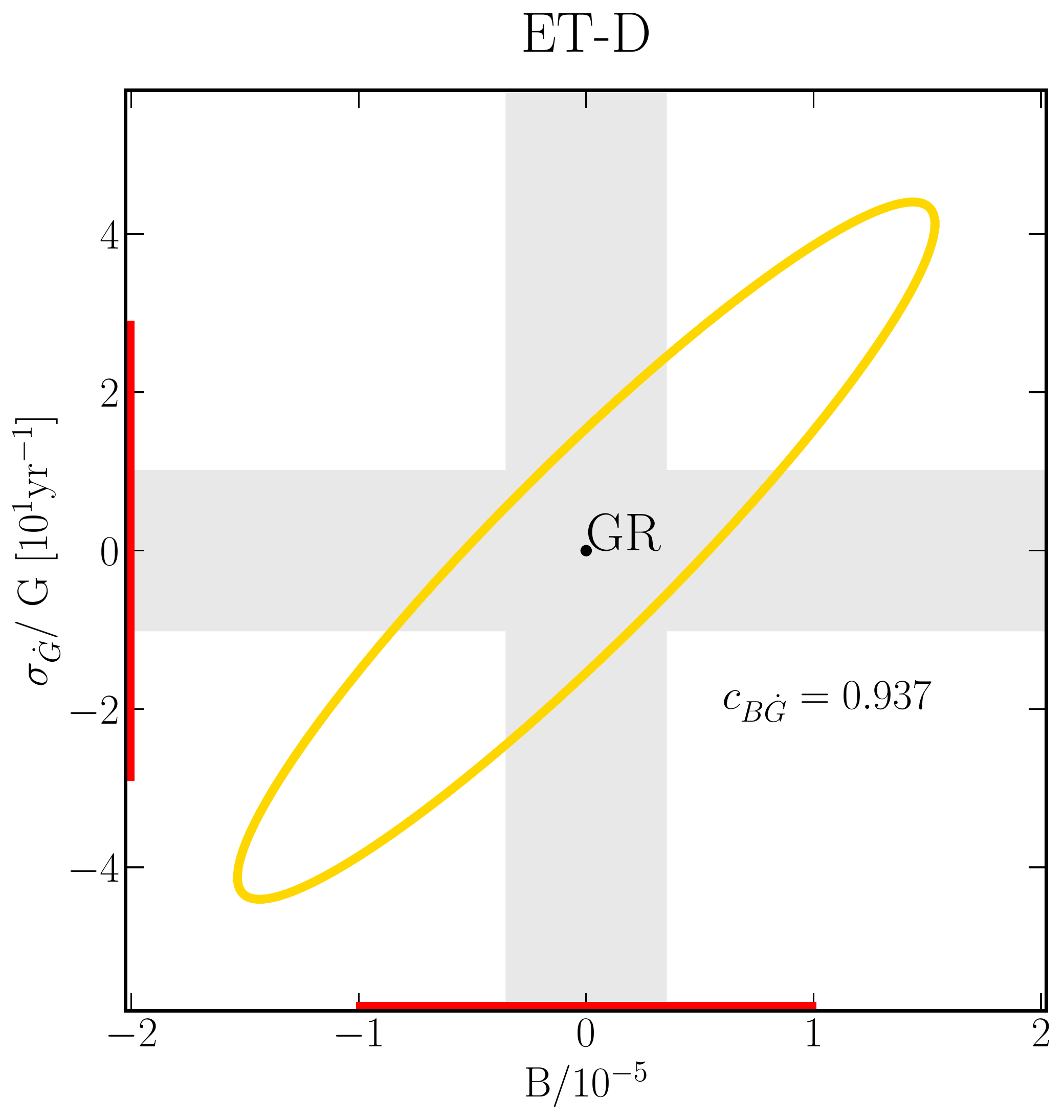}
        \label{fig:correlation_ET-D_BBH}
    \end{subfigure}
    \\
    \centering
    \begin{subfigure}[c]{0.4\textwidth}
        \includegraphics[width=0.78\textwidth]{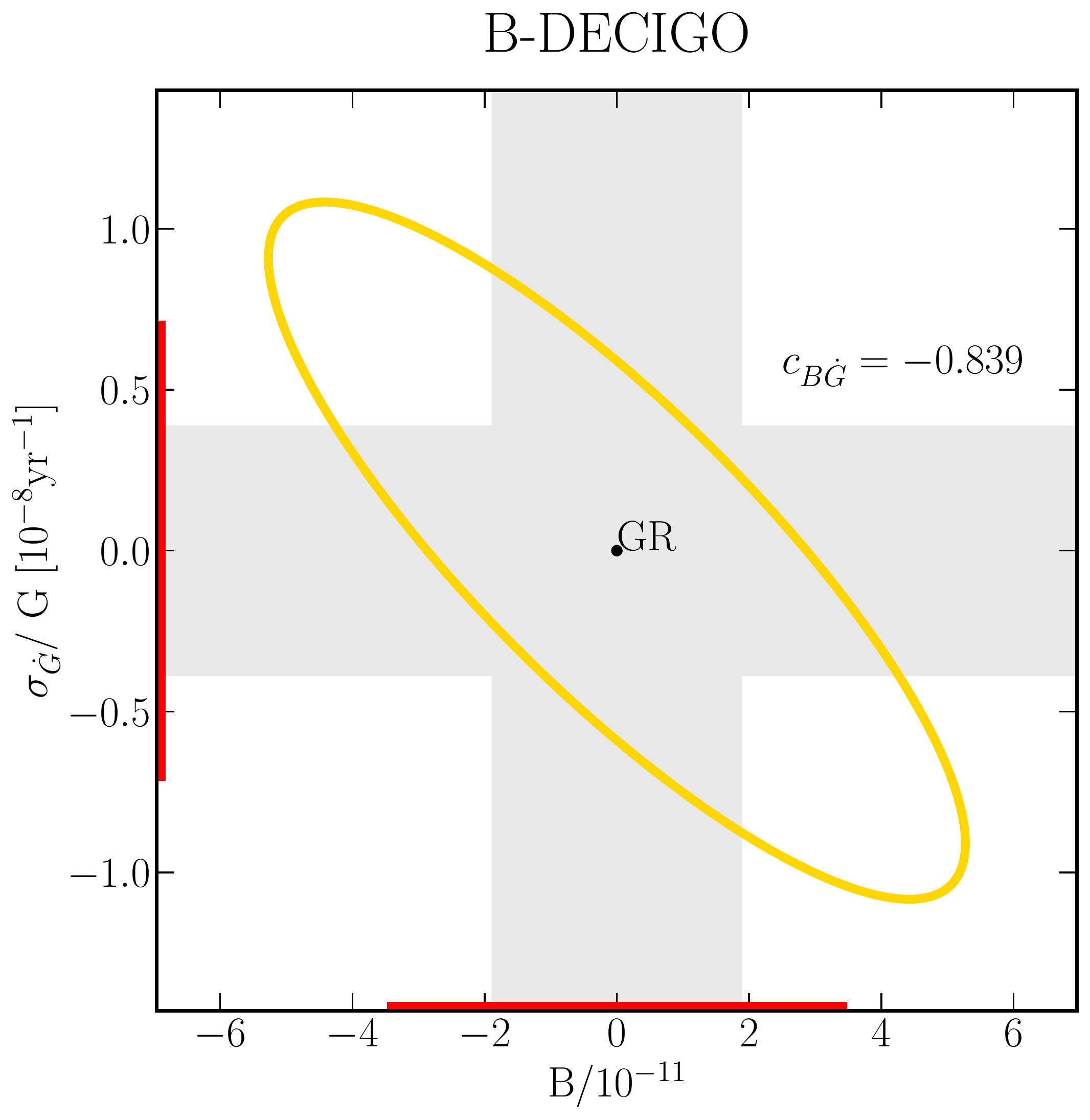}
        \label{fig:correlation_B-DECIGO_BNS}
    \end{subfigure}\hspace{1em}%
    \begin{subfigure}[c]{0.4\textwidth}
        \includegraphics[width=0.78\textwidth]{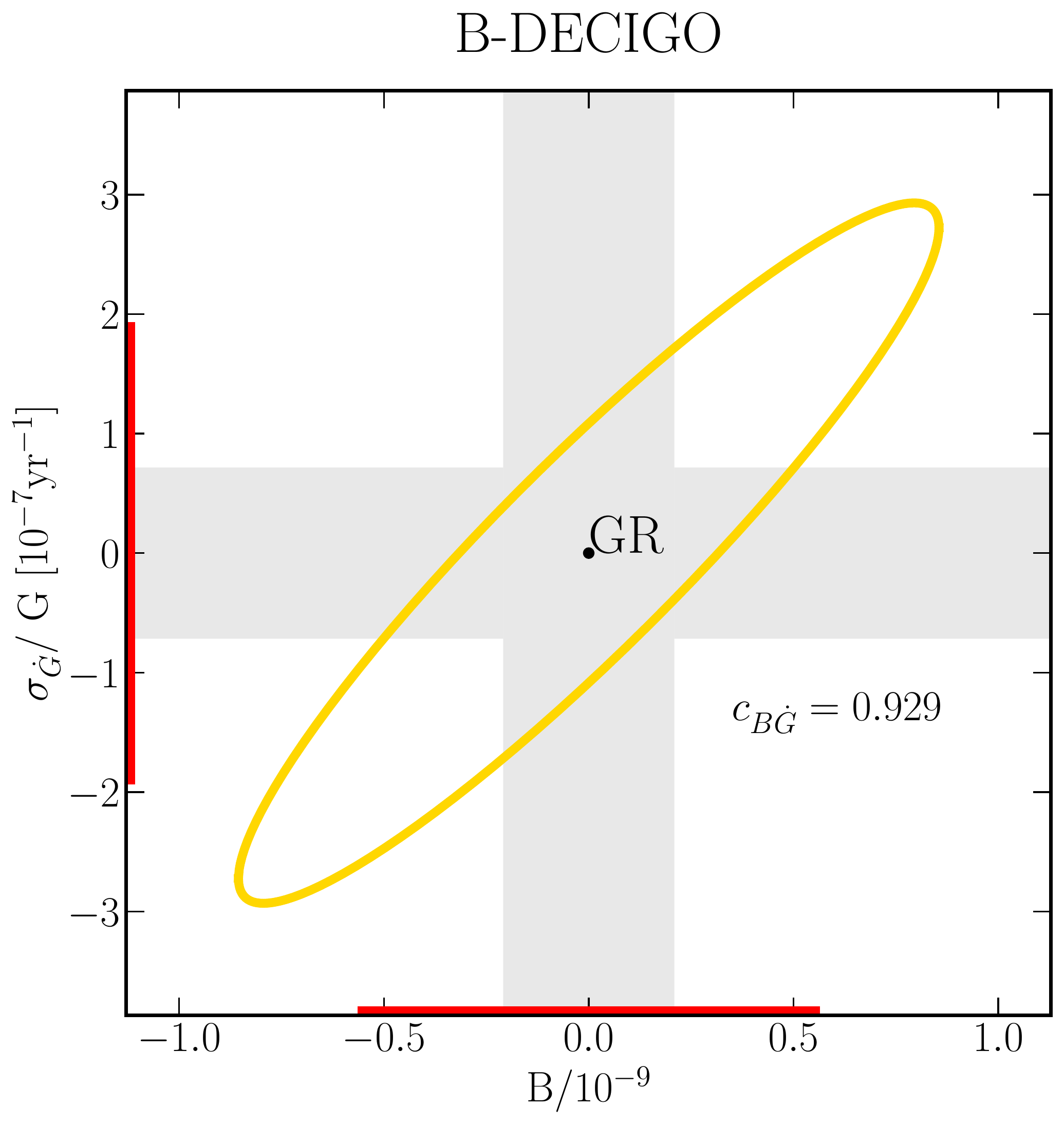}
        \label{fig:correlation_B-DECIGO_BBH}
    \end{subfigure}
    \caption{Yellow ellipses are the 1-$\sigma$ contours of the joint
    distribution of $B$ and $\Dot G$, marginalizing over other parameters. The
    standard deviations of $B$ and $\Dot G$ are indicated by the red error bars.
    Grey bands stand for the standard deviations of $B$ and $\Dot G$ when one of
    them is ignored. The left two panels are for the  GW170817-like event, while
    the right panels are for the GW150914-like event.}
    \label{fig:correlation}
\end{figure}

In order to acquire the visualized insight of the correlation between $B$ and
$\Dot G$, we use two types of detectors, ET-D and B-DECIGO, and two typical GW
events, GW150914-like and GW170817-like, as examples. In \cref{fig:correlation}
we plot the $68\%$ confidence levels of the joint posterior distributions of $B$
and $\Dot G$ using the FIM method.  The 2-dimensional Gaussian distribution of
$B$ and $\dot{G}$ can be obtained from the FIM,
\begin{align}
    P(B, \dot{G})=\frac{1}{2 \pi \sigma_{B} \sigma_{\dot{G}} \sqrt{1-c_{B
    \dot{G}}^{2}}} \exp \left\{-\frac{1}{2\left(1-c_{B
    \dot{G}}^{2}\right)}\left[\frac{B^{2}}{\sigma_{B}^{2}}-2 c_{B \dot{G}}
    \frac{B\,\dot{G}}{\sigma_{B}
    \sigma_{\dot{G}}}+\frac{\dot{G}^{2}}{\sigma_{\dot{G}}^{2}} \right]\right\}
    \,.  \label{eqn:2dposterior}
\end{align}

To make a comparison with the situation where $B$ or $\Dot G$ is overlooked, we
re-apply the FIM method to the parameter sets with $\{\ln \mathcal{A}, \ln \eta,
\ln \mathcal{M}, t_{c}, \Phi_{c}, \dot G\}$ and $\{\ln \mathcal{A}, \ln \eta,
\ln \mathcal{M}, t_{c}, \Phi_{c}, B\}$ to calculate the bound on $\Dot G$ and
$B$ alone, keeping the same values of the other parameters. We
represent these {\it individual} bounds by grey bands in~\cref{fig:correlation}.
As shown in the figure, the widths of the individual
bounds are two to three times narrower than the lengths of the corresponding
simultaneous bounds, meaning that the individual bounds turn out to
be optimistic in the parameter estimations. Calculation proves that the
ratio of simultaneous and individual bounds is $(1-c_{B
\dot{G}}^{2})^{-1/2}$. So the more that $\big| c_{B \dot{G}} \big|$ approaches
1, the stronger the correlation between $B$ and $\dot{G}$ is, and the less
authentic the individual bounds are. In \cref{fig:correlation}, $\big| c_{B\dot
G}\big|$ is between $0.84$ and $0.94$, and the simultaneous bounds on these two
parameters are about $2$ to $3$ times of the individual bounds. Therefore,
simultaneous analysis of $B$ and $\dot{G}$ is important for more precise
results. 

%---------------------------------------------------------------------
%---------------------------------------------------------------------

%---------------------------------------------------------------------
\section{Discussions}
\label{sec:dis}
%---------------------------------------------------------------------

GW has become a promising platform for GR tests. In this work, we consider two
important phenomena of violating GR, namely DGR and a varying $G$. These two
effects can occur simultaneously in some alternatives to GR. Though occurring at
different PN orders in the GW phase, they end up strongly correlated in GW
analysis. We show the simultaneous effects of constraining DGR and varying-$G$
effects using compact binary inspirals to be detected by present and future GW
detectors. We focus on two kinds of GW detectors, space-based and ground-based
laser interferometers. Using FIM, we investigate simultaneous bounds on $B$ and
$\dot{G}$ from inspiraling BNSs and BBHs.   The main results are summarized as
follow.
%--
\begin{enumerate}[(I)]
    \item Benefiting from the relatively lower accessible
    frequencies, space-based detectors make optimal output for constraining
    varying-$G$ and DGR effects. 
    %--
    \item Generally, BNSs can provide tighter bounds
    than BBHs on both $B$ and $\dot{G}$. This is particularly evident for
    ground-spaced detectors. However, limits from BBH inspirals are useful in
    some alternatives to GR, where DGR predominantly (or only) comes from
    systems involving BHs~\cite{Barausse:2016eii}, while BNSs or pulsars cannot
    give any constraints.
    %--
    \item We compare the bounds on $B$ and $\dot {G}$ from BNSs with those from
    pulsars. Among the detectors we investigated, B-DECIGO can obtain the best
    constraints with GW170817-like BNSs, reaching $B<3.4 \times 10^{-11}$ and
    $\sigma_{\Dot G}/G < 7.1 \times10^{-9} \, {\rm yr}^{-1}$ (68\% CL). The
    bound on $B$ is about 4 orders of magnitude tighter than pulsar results
    \cite{Zhu:2018etc}. For $\dot{G}$, bounds from all detectors are looser than
    pulsar results, because the effects of $\dot{G}$ are more significant in the
    lower frequency stage of inspiral. Still, the BNS inspiral signals provide
    limits in a highly dynamical strong-field regime, which is complementary to
    pulsars in a quasi-stationary strong-field regime. We look forward to
    testing varying-$G$ and DGR effects in the highly dynamical strong-field
    regime with BNS GW observations.
    %--
    \item  We also examine the correlation of $B$ and $\dot{G}$ to find out the
    necessity of a joint analysis for some alternative gravity theories. The
    correlation between them is generally strong, as indicated by $\big|
    c_{B\dot G} \big| \gtrsim 0.8$  in most detectors for  GW150914-like and
    GW170817-like events. This makes the individual bounds two to three times
    more optimistic than simultaneous bounds. So we suggest to carry out
    simultaneous estimations on both effects in future GW observations.
\end{enumerate}
%--

Our investigation can be improved by including the population distribution of
binary systems to obtain a more realistic study. Considering the binary
population can give a weighted average of constraints on $B$ and $\dot{G}$,
which would be more practical and meaningful. Also, for space-based detectors we
have used the conservative configurations of sensitivities, and better
sensitivities might be available when detectors start operating in the future.
Finally, in some regions of the parameter space the FIMs become singular, so in
this case we require further treatment to obtain credible constraints, for
example, by including the merger part of waveform. 

%---------------------------------------------------------------------
\section*{Acknowledgements}
We thank Chang Liu for useful discussions.  This work was supported by the
National Natural Science Foundation of China (11975027, 12147177, 11991053,
11721303), the National SKA Program of China (2020SKA0120300), the Max Planck
Partner Group Program funded by the Max Planck Society, and the High-performance
Computing Platform of Peking University.  ZW is supported by the Hui-Chun Chin
and Tsung-Dao Lee Chinese Undergraduate Research Endowment (Chun-Tsung
Endowment) at Peking University. JZ is supported by the ``LiYun'' Postdoctoral
Fellowship of Beijing Normal University. ZA is supported by the Principal's Fund
for the Undergraduate Student Research Study at Peking University. 
%---------------------------------------------------------------------

%---------------------------------------------------------------------
\appendix
\section{Equations used in FIM calculation}
\label{sec:appendix}
%---------------------------------------------------------------------

\allowdisplaybreaks  % allow long equations to display on different pages

Partial derivatives of the waveform used in FIM calculation are listed below. 
\begin{align} 
    \frac{\partial \tilde{h}(f)}{\partial \ln \mathcal{A}} &=\tilde{h}(f) \,, \\ 
    %--
    \frac{\partial \tilde{h}(f)}{\partial \ln \eta} &= \frac{\ii}{\eta} \fu^{-5
    / 3}\left\{ - \frac{1025}{851968} \frac{\dot G \mathcal{M} \Delta S}{\sqrt{1
    - 4 \eta} \eta^{3/5}} \fu^{-8/3} -\frac{3}{560} B \fu^{-2 / 3} 
    +\left(-\frac{743}{16128}+\frac{11}{128} \eta\right) \fu^{2 / 3}\right.
    \nonumber\\
    &+ \left.\frac{9}{40} \pi \fu+\left(-\frac{3058673}{5419008}+\frac{5429}{21504}
    \eta +\frac{617}{512} \eta^{2}\right) \fu^{4 / 3}
    +\pi\left(-\frac{7729}{4032}-\frac{38645}{32256} \ln \fu+\frac{13}{128}
    \eta\right) \fu^{5 / 3}\right.\nonumber\\ 
    &+\left[-\frac{11328104339891}{166905446400}+6 \pi^{2}+\frac{321}{35} \gE
    +\frac{107}{35} \ln (64 \fu)\right.
    +\left(\frac{3147553127}{130056192}-\frac{451}{512} \pi^{2}\right)
    \eta\nonumber\\ &\left.\left.+\frac{15211}{18432}
    \eta^{2}-\frac{25565}{6144} \eta^{3}\right] \fu^{2}
    +\pi\left(-\frac{15419335}{1548288}-\frac{75703}{32256} \eta -
    \frac{14809}{10752} \eta^2 \right) \fu^{7 / 3}\right\} \tilde{h}(f) \,, \\
    %--
    \frac{\partial \tilde{h}(f)}{\partial \ln \mathcal{M}} &= \frac{\ii}{\eta}
    \fu^{-5/3}\left\{-\frac{5}{128} +\frac{{\dot G} \mathcal{M} }{\eta^{8/5}}
    \left( \frac{1375}{1277952} - \frac{4375}{2555904} \mathcal{S} -
    \frac{5125}{2555904} \sqrt{1-4 \eta} \Delta S  \right) \fu^{-8/3} +
    \frac{1}{32} B \fu^{-2 / 3} \right. \nonumber\\
    &+\left(-\frac{3715}{32256}-\frac{55}{384} \eta\right) \fu^{2 /
    3}+\frac{\pi}{4} \fu +\left(-\frac{15293365}{65028096} -\frac{27145}{64512}
    \eta - \frac{3085}{9216} \eta^2 \right) \fu^{4 / 3}
    +\pi\left(\frac{38645}{32256}-\frac{65}{384} \eta\right) \fu^{5 / 3}
    \nonumber\\ 
    &+\left[\frac{10052469856691}{600859607040} -\frac{5}{3}
    \pi^{2}-\frac{107}{42} \gamma_{\mathrm{E}}-\frac{107}{126} \ln (64 \fu)
    +\left(-\frac{15737765635}{390168576}+\frac{2255}{1536} \pi^{2}\right)
    \eta\right.\nonumber\\ 
    &\left.\left.+\frac{76055}{221184} \eta^{2}-\frac{127825}{165888}
    \eta^{3}\right] \fu^{2}
    +\pi\left(\frac{77096675}{16257024}+\frac{378515}{96768}
    \eta-\frac{74045}{48384} \eta^{2}\right) \fu^{7 / 3}\right\} \tilde{h}(f)
    \,, \\
    %--
    \frac{\partial \tilde{h}(f)}{\partial t_{c}} &= \ii 2 \pi f \tilde{h}(f) \,,
    \\ 
    %--
    \frac{\partial \tilde{h}(f)}{\partial \Phi_{c}} &=- \ii \tilde{h}(f)\,, \\ 
    %--
    \frac{\partial \tilde{h}(f)}{\partial B} &=- \ii \frac{3}{224 \eta} \fu^{-7
    / 3} \tilde{h}(f)  \,, \\
    %--
    \frac{\partial \tilde{h}(f)}{\partial {\dot G} } &= - \ii \frac{25
    \mathcal{M}}{851968 \eta^{13/5}} \left( 11 - \frac{35}{2} S - \frac{41}{2}
    \sqrt{1 - 4 \eta} \Delta S  \right) \fu^{-13 / 3} \tilde{h}(f)   \,.
\end{align}
\bibliographystyle{apsrev}
\bibliography{refs.bib}

\end{document}